\documentclass[aps,showpacs,pra,superscriptaddress,preprintnumbers,amsmath,amssymb,twocolumn]{revtex4-2}
\usepackage[utf8]{inputenc}
\usepackage[T1]{fontenc}
\usepackage{graphicx}% Include figure files
\usepackage{dcolumn}% Align table columns on decimal point
\usepackage{appendix}
\usepackage{amsmath}
\usepackage{bm}% bold math
\usepackage[dvipsnames]{xcolor}
\usepackage{multirow}
\usepackage{notes2bib}
\usepackage{subfigure}
\usepackage{bbm}
\usepackage{enumerate}

\usepackage[
bookmarks=true,
colorlinks,
linkcolor=black,
urlcolor=black,
citecolor=black,
plainpages=false,
pdfpagelabels,
final,
breaklinks=true
]{hyperref}

\usepackage{titlesec}

\usepackage{array}

\usepackage{physics}

\begin{document}
	\allowdisplaybreaks
	
	\title{Squeezed states of light after high-harmonic generation in excited atomic systems}
	
	\author{J.~Rivera-Dean}
	\email{javier.rivera@icfo.eu}
	\affiliation{ICFO -- Institut de Ciencies Fotoniques, The Barcelona Institute of Science and Technology, 08860 Castelldefels (Barcelona)}
	
	\author{H. B. Crispin}
	\affiliation{Physics Program, Guangdong Technion--Israel Institute of Technology, Shantou, Guangdong 515063, China}
	\affiliation{Guangdong Provincial Key Laboratory of Materials and Technologies for Energy Conversion, Guangdong Technion – Israel Institute of Technology, Shantou, Guangdong 515063, China}
	\affiliation{Department of Physics, Technion – Israel Institute of Technology, Technion City, Haifa, 32000, Israel}
	\affiliation{Solid State Institute, Technion – Israel Institute of Technology, Technion City, Haifa, 32000, Israel}
	\affiliation{Helen Diller Quantum Center, Technion – Israel Institute of Technology, Technion City, Haifa, 32000, Israel}
	
	\author{P.~Stammer}
	\affiliation{ICFO -- Institut de Ciencies Fotoniques, The Barcelona Institute of Science and Technology, 08860 Castelldefels (Barcelona)}
	\affiliation{Atominstitut, Technische Universität Wien, 1020 Vienna, Austria}
	
	\author{Th. Lamprou}
	\affiliation{Foundation for Research and Technology-Hellas, Institute of Electronic Structure \& Laser, GR-70013 Heraklion (Crete), Greece}
	
	\author{E.~Pisanty}
	\affiliation{Attosecond Quantum Physics Laboratory, Department of Physics, King's College London, Strand WC2R 2LS, London, United Kingdom}
	
	\author{M.~Krüger}
	\affiliation{Department of Physics, Technion – Israel Institute of Technology, Technion City, Haifa, 32000, Israel}
	\affiliation{Solid State Institute, Technion – Israel Institute of Technology, Technion City, Haifa, 32000, Israel}
	\affiliation{Helen Diller Quantum Center, Technion – Israel Institute of Technology, Technion City, Haifa, 32000, Israel}
	
	\author{P. Tzallas}
	\affiliation{Foundation for Research and Technology-Hellas, Institute of Electronic Structure \& Laser, GR-70013 Heraklion (Crete), Greece}
	\affiliation{ELI-ALPS, ELI-Hu Non-Profit Ltd., Dugonics tér 13, H-6720 Szeged, Hungary}
	
	\author{M. Lewenstein}
	\affiliation{ICFO -- Institut de Ciencies Fotoniques, The Barcelona Institute of Science and Technology, 08860 Castelldefels (Barcelona)}
	\affiliation{ICREA, Pg. Llu\'{\i}s Companys 23, 08010 Barcelona, Spain}
	
	\author{M.~F.~Ciappina}
	\email{marcelo.ciappina@gtiit.edu.cn}
	\affiliation{Physics Program, Guangdong Technion--Israel Institute of Technology, Shantou, Guangdong 515063, China}
	\affiliation{Guangdong Provincial Key Laboratory of Materials and Technologies for Energy Conversion, Guangdong Technion – Israel Institute of Technology, Shantou, Guangdong 515063, China}
	\affiliation{Technion -- Israel Institute of Technology, Haifa, 32000, Israel}

	\date{\today}
	\begin{abstract}
		High-harmonic generation (HHG) has recently emerged as a promising method for generating non-classical states of light with frequencies spanning from the infrared up to the extreme ultraviolet regime.~In this work, we theoretically investigate the generation of squeezed states of light through HHG processes in atomic systems that had been initially driven to their first excited state.~Our study reveals significant single-mode squeezing in both the driving field and low-order harmonic modes.~Additionally, we characterize two-mode squeezing features in the generated states, both between fundamental and harmonic modes, and among the harmonic modes themselves.~Using these correlations, we demonstrate the generation of optical Schrödinger kitten states through heralding measurements, specifically via photon subtraction in one of the modes influenced by two-mode~squeezing.
	\end{abstract}
	\maketitle
	
	\section{INTRODUCTION}
	Non-classical states of light are defined as states whose properties cannot be described by classical electromagnetism, and therefore require the framework of quantum optics.~These states have become crucial for the development of photonic quantum technologies~\cite{Benyoucef_book_ch2}, providing versatile and resilient tools for quantum information science applications~\cite{gisin_quantum_2007,kok_linear_2007,giovannetti_advances_2011,aspuru-guzik_photonic_2012,pirandola_advances_2018}.~Among them, squeezed states have been of fundamental importance.~These states exhibit reduced levels of noise for specific physical observables compared to classical states, at the expense of increased uncertainty in the corresponding conjugate observable, while still complying with Heisenberg's uncertainty principle~\cite{walls_squeezed_1983,ScullyBookCh2,GerryBookCh7}.~For this reason, they have been essential for improving measurement sensitivity in interferometric setups~\cite{caves_quantum-mechanical_1981,slusher_observation_1985,xiao_precision_1987,jaekel_quantum_1990,kimble_conversion_2001,mckenzie_experimental_2002,eberle_quantum_2010,abadie_gravitational_2011}.~Additionally, they have been used as a source of more elaborate non-classical states of light~\cite{dakna_generating_1997,ourjoumtsev_generating_2006,neergaard-nielsen_generation_2006}.
	
	The generation of squeezed states in practice typically relies on nonlinear optical processes such as four-wave mixing~\cite{slusher_observation_1985}, optical parametric oscillators~\cite{wu_squeezed_1987,lam_optimization_1999}, or optical parametric amplification~\cite{schneider_generation_1998}.~However, despite high-harmonic generation (HHG) being one of the most prominent examples of nonlinear optical phenomena known to date, its capabilities for generating squeezed states and other non-classical states of light have only recently been explored~\cite{bhattacharya_stronglaserfield_2023,lewenstein_attosecond_2024}.~In HHG, high-intensity infrared (IR) radiation is upconverted into high-frequency radiation after subfemtosecond electron dynamics driven in a given matter system, producing frequencies that span from the infrared regime to the extreme ultraviolet (XUV)~\cite{burnett_harmonic_1977,mcpherson_studies_1987,ferray_multiple-harmonic_1988}.~Consequently, HHG has been used in various applications: from the generation of XUV radiation sources~\cite{drescher_x-ray_2001,silva_spatiotemporal_2015} to attosecond science~\cite{corkum_attosecond_2007,krausz_attosecond_2009}.~Within a quantum optical context, recent experimental~\cite{lewenstein_generation_2021,rivera-dean_strong_2022,bhattacharya_stronglaserfield_2023,lamprou_nonlinear_2023} and theoretical~\cite{rivera_dean_quantum-optical_2019,gorlach_quantum-optical_2020,stammer_high_2022,stammer_theory_2022,stammer_quantum_2023} studies have shown that HHG can produce states of light with seemingly non-classical features, including squeezed states, as evidenced in experiments~\cite{theidel_evidence_2024} and theoretical models~\cite{stammer_squeezing_2023,lange_electron-correlation-induced_2024,yi_generation_2024}. Furthermore, squeezed states have been used to drive HHG processes both experimentally~\cite{rasputnyi_high_2024,lemieux_photon_2024} and theoretically~\cite{gorlach_high-harmonic_2023,even_tzur_photon-statistics_2023}, with predictions suggesting the generation of high-order harmonic squeezed states~\cite{tzur_generation_2023}.
	
	The generation of squeezed states and other non-classical states of light directly from classical HHG setups, where a classical driving field interacts with a matter system, strongly depends on the cross-talk between different energy levels of the system.~In Ref.~\cite{stammer_squeezing_2023}, the observed squeezing and entanglement features resulted from correlations between the time-dependent dipole moment at different times.~These correlations become significant when fields with sufficient intensity are used, which inevitably causes non-negligible depletion of the ground state of the system~\cite{delone_tunneling_1998,tong_empirical_2005}.~Under these conditions, substantial squeezing was observed in the fundamental mode, with properties depending on the specific excitation conditions.~In Ref.~\cite{lange_electron-correlation-induced_2024}, a theoretical analysis of HHG driven in a 1D Fermi-Hubbard model revealed significant non-classical features on the field when the matter system was in the Mott-insulating phase, where the coupling between different sites in the chain becomes non-vanishing.~Conversely, these features were absent in the corresponding uncorrelated phase.~Similar results were observed in the context of HHG in semiconductor materials~\cite{rivera-dean_quantum_2023}, attributed to the delocalized nature of the electron dynamics resulting in HHG~\cite{osika_wannier-bloch_2017,parks_wannier_2020}.~Furthermore, in solid-state systems, experimental observations have indicated the presence of two-mode squeezing features between low-order harmonics~\cite{theidel_evidence_2024}, which seem to result from Bloch oscillations of electrons within the conduction band of the solid~\cite{gonoskov_nonclassical_2024}. 
	
	In this work, we deviate from these studies by considering the active participation of excited states of atomic systems during the HHG process~\cite{watson_harmonic_1996,sanpera_harmonic-generation_1996,paul_enhanced_2005}.~Recently, Ref.~\cite{yi_generation_2024} demonstrated that the backaction of generated harmonics on the coupling between ground and excited states, potentially enhanced by a cavity, can be a versatile tool for generating various non-classical states of light.~This includes squeezed-like states, with properties strongly influenced by the mean photon number of the generated harmonics.~Additionally, Ref.~\cite{rivera-dean_quantum-optical_2024} reported non-classical features affecting both light and matter when driving HHG in diatomic molecular systems.~These features arise from the involvement of both ground and excited states due to the various pathways electrons can take during HHG~\cite{faria_high-order_2007,suarez_high-order-harmonic_2017,suarez_rojas_strong-field_2018}.~Moreover, Ref.~\cite{pizzi_light_2023} observed that initial entanglement features involving the ground and excited states of a many-body system of atoms could be mapped onto the generated harmonics.
	
	Here, we explore a setup in which atomic systems are initially pumped to their first excited state before interacting with a classical, strong-laser field~\cite{watson_harmonic_1996,sanpera_harmonic-generation_1996}.~The pumping process can be achieved by using a pulse resonant with the transition frequency prior to the strong-field interaction, as in similar experimental implementations within the strong-field regime~\cite{paul_enhanced_2005}.~Within this framework, we report squeezing features on the quantum optical state that emerge from the cross-talk between the electronic ground and first excited states mediated by the time-dependent dipole moment.~Notably, these squeezing features appear only when the electron initiates and ends the dynamics in the first excited state, and they are absent in those instances when the electron recombines with the ground state.~We observe that these squeezing effects influence both the input driving field and the generated harmonic modes under field parameters typical of standard HHG experiments.~Additionally, we observe significant two-mode squeezing features between different field modes.~We demonstrate that these correlations are sufficient to generate other types of non-classical states of light, such as optical Schrödinger kitten states, via heralding measurements, specifically through photon substraction in one of the modes affected by two-mode squeezing.~This approach offers an alternative to the postselection schemes described in Refs.~\cite{lewenstein_generation_2021,rivera-dean_strong_2022,stammer_high_2022,stammer_theory_2022,stammer_quantum_2023} for producing optical Schrödinger kitten-like states via HHG processes.
	%~With this work, we aim to provide novel pathways for generating bright single-mode and two-mode squeezed states through HHG.
	
	\section{THEORETICAL BACKGROUND}
	Our analysis is based on solving the time-dependent Schrödinger equation describing the interaction of an atomic system with a quantized field, assuming the initial state of the system is $\ket{\Psi(t_0)} = \ket{\text{e}} \otimes \ket{\alpha_L} \bigotimes_{q=2}^{q_c} \ket{0_q}$. This implies that the electron is initially in the (non-degenerate) first excited state of the atom ($\ket{\text{e}}$), and the field is in a product of coherent states where mode $q=1\equiv L$ has amplitude $|\alpha_L| \gg 1$, with the rest in the vacuum state. This scenario can be prepared by first applying a $\pi$-pulse resonant with the transition between the atomic ground ($\ket{\text{g}}$) and first excited states, followed by a strong-laser field represented by the coherent state $\ket{\alpha_L}$. Under the length gauge, and the dipole and single-active-electron approximations, the Schrödinger equation for this system can be expressed as~\cite{lewenstein_generation_2021,rivera-dean_strong_2022,stammer_quantum_2023}
	\begin{equation}\label{Eq:TDSE}
		i\hbar \pdv{\ket{\Psi(t)}}{t}
		= \hat{E}(t) \hat{d}(t) \ket{\Psi(t)},
	\end{equation}
	in the interaction picture with respect to the free field Hamiltonian $\hat{H}_{\text{field}} = \sum_{q=1}^{q_c} \hbar \omega_q \hat{a}^\dagger_q \hat{a}_q$, in the displaced framework with respect to $\ket{\alpha_L}$ and within the interaction picture with respect to the semiclassical Hamiltonian $\hat{H}_{\text{sc}}(t) = \hat{H}_{\text{at}} + E_{\text{cl}}(t) \hat{d}$.~Here, $\hat{d}$ represents the dipole moment operator and $\hat{d}(t) = \hat{U}_{\text{sc}}(t) \hat{d} U^\dagger_{\text{sc}}(t)$ its time-dependent version, with $\hat{U}_{\text{sc}}(t,t_0) = \hat{\mathcal{T}} \text{exp}[-\tfrac{i}{\hbar} \int^t_{t_0} \dd \tau \hat{H}_{\text{sc}}(\tau)]$, where $\hat{T}$ represents the time-ordering operator; $\hat{E}(t) = \sum_{q=1}^{q_c}\hat{E}_q(t) = -if(t) \sum^{q_c}_{q=1}g(\omega_q) [\hat{a} e^{-i\omega_q t} - \text{h.c.}]$ is the electric field operator with $0 \leq f(t) \leq 1$ an envelope function compatible with the applied field envelope and $g(\omega) \equiv \sqrt{\hbar\omega/(2V\epsilon_0)}$; and $\hat{E}_{\text{cl}}(t) =  \langle\alpha_L, \{0\}_{q=2}^{q_c}\vert \hat{E}(t) \vert \alpha_L, \{0\}_{q=2}^{q_c}\rangle$. While a fully rigorous quantum optical analysis would typically require an infinite number of modes to represent a continuous spectrum, we apply the envelope function $f(t)$ both to confine the interaction to a finite time interval and to recover the classical-pulsed field expressions. This approach yields results equivalent to those obtained when using multimode descriptions of the driving laser field~\cite{rivera-dean_strong_2022,stammer_quantum_2023}, while preserving the simplicity of using single-mode drivers which in particular benefits numerical analyses.~However, it is important to note that the expressions derived in this work are formulated in terms of $\hat{E}(t)$ and $E_{\text{cl}}(t)$, making our results readily extendable to more comprehensive multimode descriptions.
	
	Within the displaced framework of Eq.~\eqref{Eq:TDSE}, under which the initial state reads as $\ket{\Psi(t_0)} = \ket{\text{e}} \bigotimes_{q=1}^{q_c}  \ket{0_q} \equiv \ket{\text{e}} \otimes \ket{\bar{0}}$, we proceed to solve the dynamics.~To do so, similarly to Refs.~\cite{lewenstein_generation_2021,rivera-dean_strong_2022,stammer_quantum_2023}, we neglect the continuum populations at all times, assuming their contribution to be small compared to those of the lowest-energy atomic states~\cite{watson_harmonic_1996,sanpera_harmonic-generation_1996}. This leads to the following set of coupled differential equations (see Appendix \ref{App:Derivation})
	\begin{align}
		&i\hbar \pdv{\ket{\Phi_{\text{g}}(t)}}{t}
		= \mu_{\text{gg}}(t)\hat{E}(t)
		\ket{\Phi_{\text{g}}(t)}
		+ \mu_{\text{ge}}(t)\hat{E}(t)
		\ket{\Phi_{\text{e}}(t)},\label{Eq:TDSE:g}
		\\&
		i\hbar \pdv{\ket{\Phi_{\text{e}}(t)}}{t}
		= \mu_{\text{eg}}(t)\hat{E}(t)
		\ket{\Phi_{\text{g}}(t)}
		+ \mu_{\text{ee}}(t)\hat{E}(t)
		\ket{\Phi_{\text{e}}(t)},\label{Eq:TDSE:e}
	\end{align}
	where $\mu_{\text{ij}}(t) \equiv \langle \text{i} \vert \hat{d}(t)\vert \text{j}\rangle$ and $\ket{\Phi_{\text{i}}(t)} \equiv \braket{\text{i}}{\Psi(t)}$.~It is worth noting that although continuum states are not taken into account in the interaction picture we are working with, they are considered in the computation of $\mu_{\text{ij}}(t)$~\cite{watson_harmonic_1996,sanpera_harmonic-generation_1996}.
	
	For the parameters considered here (peak intensity $I_0 = 10^{14}$ W/cm$^2$, central wavelength $\lambda_L = 800$ nm, and ionization potentials  $I_{p,\text{g}} = 54.4$ eV and $I_{p,\text{e}} = 13.6$ eV), it is found that $\abs{\mu_{\text{ee}}(t)} \gg \abs{\mu_{\text{eg}}(t)} \gg \abs{\mu_{\text{gg}}(t)}$ (see Appendix~\ref{App:Derivation}). This allows us to write the following perturbation theory solution for Eq.~\eqref{Eq:TDSE:g}
	\begin{equation}\label{Eq:Solution:ground}
		\ket{\Phi_{\text{g}}(t)}
		\approx 
		-\dfrac{i}{\hbar}
		\int^t_{t_0} \dd\tau
		\mu_{\text{ge}}(\tau)\hat{E}(\tau) \ket{\Phi_{\text{e}}(\tau)},
	\end{equation}
	and introducing this expression into Eq.~\eqref{Eq:TDSE:e} results in
	\begin{equation}\label{Eq:TDSE:e:pert}
		\begin{aligned}
			i\hbar \pdv{\ket{\Phi_{\text{e}}(t)}}{t}
			&\approx
			\mu_{\text{ee}}(t) \hat{E}(t) \ket{\Phi_{\text{e}}(t)}
			+ \mu_{\text{eg}}(t) \hat{E}(t) \ket{\Phi_{\text{g}}(t)}
			\\&\quad
			- \dfrac{i}{\hbar} \mu_{\text{eg}}(t) \hat{E}(t)
			\int^t_{t_0} \dd \tau \mu_{\text{ge}}(\tau) \hat{E}(\tau) \ket{\Phi_{\text{e}}(\tau)},
		\end{aligned}
	\end{equation}
	where we note here that we have now accounted for the initial conditions specified earlier, assuming the electron is initially in the atom’s first excited state. The more general scenario, with arbitrary initial state conditions for the atomic system, is evaluated in Appendix~\ref{App:Derivation}.
	
	To simplify the analysis of Eq.~\eqref{Eq:TDSE:e:pert}, we perform a Markov-like approximation, i.e. $\ket{\Phi_{\text{e}}(\tau)} \to \ket{\Phi_{\text{e}}(t)}$, which neglects transient changes on the quantum optical state between times $t_0$ and $t$, effectively assuming that the cumulative influence of past interactions can be represented by an instantaneous operator acting on the state at time $t$.These transient changes could be related to quantum optical fluctuations arising from the electronic oscillation during its excursion in the continuum.~A more detailed analysis of these effects and their influence during the HHG processes can be found in Ref.~\cite{rivera-dean_about_2024}.~Under this approximation, a solution to Eq.~\eqref{Eq:TDSE:e:pert} can be expressed as (see Appendix~\ref{App:Derivation})
	\begin{equation}\label{Eq:Solution:excited}
		\begin{aligned}
			\ket{\Phi_{\text{e}}(t)}
			&\approx \hat{U}(t,t_0) \ket{\bar{0}}
			\\&\quad
			+ \dfrac{1}{\hbar^2}
			\int^t_{t_0} \dd \tau_1
			\hat{U}(t,\tau_1) 
			\mu_{\text{eg}}(\tau_1)
			\hat{E}(\tau_1)
			\\&\hspace{1cm}\times
			\int^{\tau_1}_{t_0} \dd \tau_2
			\mu_{\text{ge}}(\tau_2) \hat{E}(\tau_2) \hat{U}(\tau_2,t_0)\ket{\bar{0}},
		\end{aligned}
	\end{equation}
	where $\hat{U}(t,t_0)$ is given, up to a phase prefactor, by
	\begin{equation}
		\hat{U}(t,t_0)
		= \hat{\vb{D}}\big(\boldsymbol{\chi}_\text{e}(t)\big)
		\exp[-\dfrac{i}{\hbar} \int^t_{t_0} \dd \tau \hat{Q}(\tau)],
	\end{equation}
	where $\hat{\vb{D}}(\boldsymbol{\chi}_{\text{e}}) \equiv \prod_{q=1}^{q_c} \hat{D}(\chi^{(q)}_{\text{e}}(t))$ with $\hat{D}_q(\cdot)$ being the displacement operator acting on the $q$th harmonic mode~\cite{ScullyBookCh2}; $\chi^{(q)}_{\text{e}}(t) \propto g(\omega_q) \int^{t}_{t_0} \dd \tau \mu_{\text{ee}}(\tau) e^{i\omega_q \tau}$; and $\hat{Q}(t) \equiv \frac{i}{\hbar}\mu_{\text{eg}}(t) \hat{E}(t) \int^t_{t_0} \dd \tau \mu_{\text{ge}}(\tau) \hat{E}(\tau)$.
	
	From Eq.~\eqref{Eq:Solution:excited} we see that in addition to the displacement operator, the field degrees of freedom are also affected by an operator that is second order in $\hat{E}(t)$. This involves second-order terms with respect to creation and annihilation operators acting on the different modes, ultimately leading to the presence of squeezing and entanglement between the field modes \cite{stammer_squeezing_2023}. These phenomena originate from the cross-talk between the ground and first excited states through the time-dependent dipole moment operator $\hat{d}(t)$, represented here by $\mu_{\text{eg}}(t)$~\cite{rivera-dean_quantum-optical_2024}.~It is worth noting that this cross-talk might only involve population exchanges mediated by the electronic continuum states if both excited and ground states have the same parity.
	
	It is also important to note that the contribution of these squeezing-like terms is much smaller than the contributions stemming from the displacement $\chi_{\text{e}}^{(q)}(t)$, as the former are proportional to $g(\omega_L)^2$ while the latter to $g(\omega_L)$, with $g(\omega_L)$ being a perturbative quantity~\cite{rivera-dean_light-matter_2022,rivera-dean_quantum-optical_2024}. Consequently, to observe squeezing and entanglement effects, it is necessary to consider the collective contribution of many atoms~\cite{stammer_squeezing_2023}. This, a priori, involves solving the strong-field dynamics for a many-body problem involving a total of $N_{\text{at}}$ atoms, with each atom coupled to the same electromagnetic field. However, assuming that the time-dependent dipole moments of different atoms are uncorrelated~\cite{sundaram_high-order_1990}, the potential correlations between different atoms arising due to this light-mediated dynamics average to zero  (see Appendix~\ref{App:MB}), with standard deviations being proportional to $g(\omega_L)^4$.~Hence, from the point of view of our analysis, which focuses on the squeezing properties of the field that scale as $g(\omega_L)^2$, we can regard the interaction of each atom with the electromagnetic field as independent.~This implies that, assuming all atoms are initially in the same state, the many-body evolution can be approximately written as
	\begin{equation}\label{Eq:Sol:MB}
		\lvert\Psi(t)\rangle
		\approx
		\prod_{i=1}^{N_{\text{at}}}
		\vb{U}(t,t_0)
		\bigotimes_{i=1}^{N_{\text{at}}}\ket{\text{e}}
		\otimes \ket{\bar{0}},
	\end{equation}
	with $\vb{U}(t,t_0)$ the time-evolution operator obtained from Eq.~\eqref{Eq:TDSE}, and that approximately leads to Eqs.~\eqref{Eq:Solution:ground} and~\eqref{Eq:Solution:excited}.
	
	Among all possible outcomes, we particularly focus on scenarios where all electrons return to their initial state, i.e., the excited state.~Hence, upon returning to the original frame of reference for the electronic degrees of freedom, we obtain (see Appendix \ref{App:MB})
	\begin{align}
		\lvert\bar{\Phi}_{\text{e}}(t)\rangle
		&= \bigg[
		\hat{U}(t,t_0)\nonumber
		\\&\quad
		+ \dfrac{1}{\hbar^2}
		\int^t_{t_0} \dd \tau_1
		\hat{U}(t,\tau_1) 
		\mu_{\text{eg}}(\tau_1)
		\hat{E}(\tau_1)\nonumber
		\\&\hspace{0.7cm}\times
		\int^{\tau_1}_{t_0} \dd \tau_2
		\mu_{\text{ge}}(\tau_2) \hat{E}(\tau_2) \hat{U}(\tau_2,t_0)
		\bigg]^{N_{\text{at}}} \!\ket{\bar{0}}\label{Eq:Sol:QO:e:1}
		\\& \approx 
		\big[
		\hat{U}(t,t_0)
		\big]^{N_{\text{at}}}
		\ket{\bar{0}},\label{Eq:Sol:QO:e:2}
	\end{align}
	where we approximate the total state by considering only the first contribution, which is valid under the condition $N_{\text{at}}g(\omega_L)^2 \lesssim 1$, ensuring that the second term in Eq.~\eqref{Eq:Sol:QO:e:1} is significantly smaller than the first one. This condition sets the stage for the results discussed in the next section.
	
	It is worth noting here that when events where the electron recombines in the ground state, or begins in the ground state and recombines in the excited state, are considered, we observe delocalized single-photon excitations spanning multiple harmonic modes (see also Appendix~\ref{App:Derivation}).~This configuration leads to non-classicality and entanglement across harmonic modes, similar to those observed in the quantum optical treatment of HHG in molecular systems in Ref.~\cite{rivera-dean_quantum-optical_2024}.~Due to these similarities, and given our focus on squeezing features, we refer interested readers to the aforementioned reference for further details.
	
	\section{RESULTS}
	In this section, we characterize the squeezing properties of the state in Eq.~\eqref{Eq:Sol:QO:e:2} using a constructive approach.~Initially, we neglect the contribution of $\hat{a}_q \hat{a}_{q'}$ and related second-order terms where $q \neq q'$, to identify which harmonic modes are most affected by squeezing.~Subsequently, we incorporate these additional terms. The numerical analysis presented in this section considers an applied pulse with sin$^2$ envelope, $I_0 = 10^{14}$ W/cm$^2$, $\lambda_L = 800$ nm and $\Delta t \approx 16$ fs of duration, while setting the ionization potentials of the atomic system to $I_{p,\text{g}} = 54.4$ eV and $I_{p,\text{e}} = 13.6$ eV, corresponding to those of He$^+$~\cite{watson_harmonic_1996,sanpera_harmonic-generation_1996}.~Details about the numerical implementation can be found along the Appendix. While our theory is sufficiently general to encompass atomic systems that satisfy the $\lvert\mu_{\text{gg}}(t)\rvert < \lvert\mu_{\text{eg}}(t)\rvert$ condition, our selection of He$^+$ is based on its lack of significant HHG radiation when the electron begins in the ground state, thereby meeting the aforementioned condition. Additionally, its prior use in semiclassical analyses provides a valuable benchmark for our results~\cite{watson_harmonic_1996,sanpera_harmonic-generation_1996}. In fact, experimental observations of HHG using initial electronic superpositions spanning several excited states have involved alkali atoms, with a near-infrared driver to create the initial excited state and a mid-infrared driver for the strong-field interaction~\cite{paul_enhanced_2005}.
	
	\subsection{Single-mode squeezing}
	When referring to single-mode squeezed states, we are describing quantum optical states where the uncertainty in one optical quadrature, hereupon $\Delta\hat{X} \equiv \langle \hat{X}^2\rangle - \langle \hat{X}\rangle^2$, is reduced below the uncertainty found for coherent states ($\Delta X = 1/2$), at the expense of increased uncertainty in the conjugate quadrature $\Delta \hat{\bar{X}}$.~Here, we aim to determine whether states in the form of Eq.~\eqref{Eq:Sol:QO:e:2}, after neglecting correlations between the different optical modes, can exhibit such features.~This expectation arises since, in that scenario, Eq.~\eqref{Eq:Sol:QO:e:2} can be rewritten as
	\begin{equation}\label{Eq:single:mode:approx}
		\lvert \bar{\Phi}_{\text{e}}(t)\rangle
		\approx \hat{\vb{D}}\big(\Tilde{\boldsymbol{\chi}}\big)
		\exp[-\sum_{q=1}^{q_c}
		\dfrac{iN_{\text{at}}}{\hbar}
		\int^t_{t_0}
		\dd \tau \hat{Q}_{q,q}(\tau)] \ket{\bar{0}},
	\end{equation}
	where $\hat{Q}_{q,q}(t) \equiv \frac{i}{\hbar}\mu_{\text{eg}}(t) \hat{E}_q(t) \int^t_{t_0} \dd \tau \mu_{\text{ge}}(\tau) \hat{E}_q(\tau)$, which involves second-order terms of creation and annihilation operators. In this expression, $\Tilde{\boldsymbol{\chi}} \equiv f(\boldsymbol{\chi}_{\text{e}}(t))$, where $f(\cdot)$ is a function that depends on the prefactors accompanying the product between creation and annihilation operators in $\hat{Q}_{q,q}$~\footnote{This can be easily seen from the fact that we can always write $[f(\hat{a}^2, \hat{a}^\dagger\hat{a}), g(\hat{a})] = h(\hat{a})$, with $f(\cdot)$, $g(\cdot)$ and $h(\cdot)$ linear functions of the corresponding operators. Therefore, $\exp[-f(\hat{a}^2, \hat{a}^\dagger\hat{a})]\exp[g(\hat{a})] \exp[f(\hat{a}^2, \hat{a}^\dagger\hat{a})] = \exp[\Tilde{h}(\hat{a})]$, where $\Tilde{h}(\cdot)$ does not necessarily have to be equal to $h(\cdot)$}.
	
	\begin{figure}
		\centering
		\includegraphics[width=1\columnwidth]{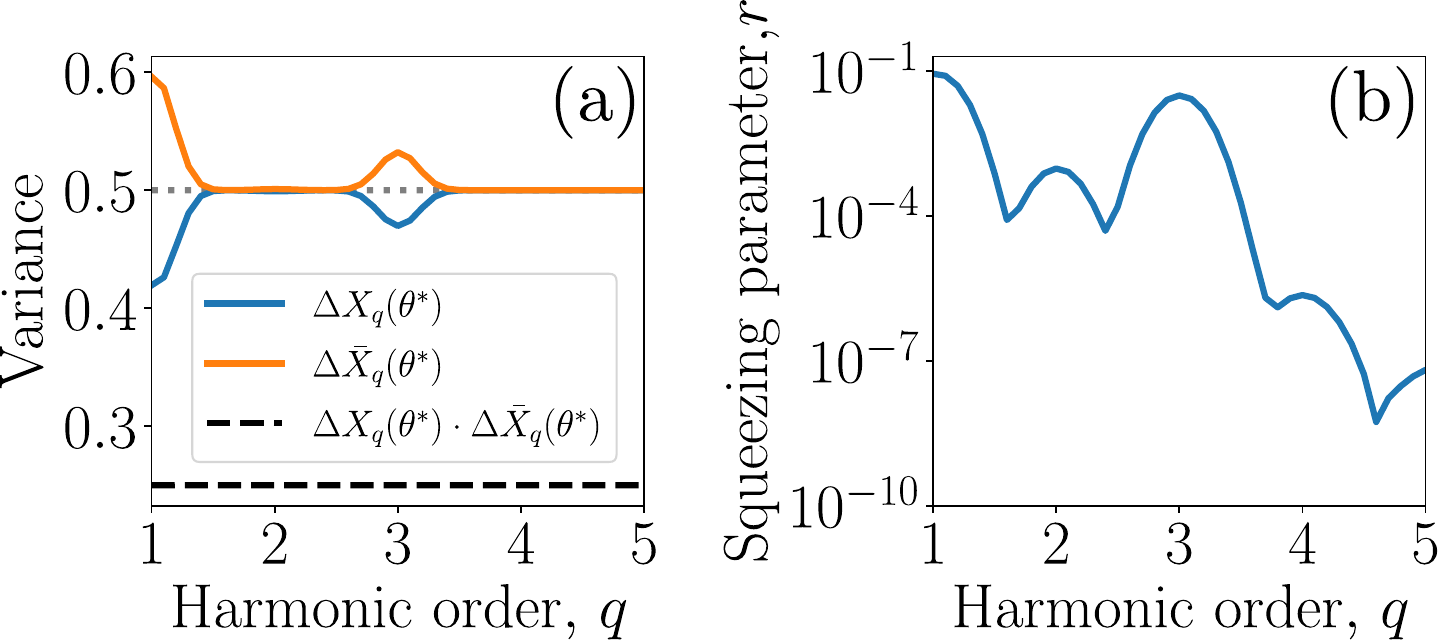}
		\caption{Amount of single-mode squeezing for different harmonic orders.~In panel (a), the blue curve shows the results of the optimization from Eq.~\eqref{Eq:minimization:single}, while the orange curve displays the variance of the conjugate variable.~Their product is represented by the dashed black curve, corresponding to the minimum Heisenberg uncertainty.~The gray dotted line shows $\Delta X_q = \Delta \bar{X}_q$ for the case of having a coherent state in each harmonic mode. In panel (b), the squeezing parameter is computed as $r = -(1/2) \log_{10}(2 \Delta X_q(\theta^*))$.~We set $N_{\text{at}}g(\omega_L)^2=1$ a.u., with the atomic units (a.u.) obtained by setting $\hbar = m_{\mathsf{e}} = \abs{\mathsf{e}} = 1$.}
		\label{Fig:Amount:Squeezing}
	\end{figure}
	
	Given the definition of squeezed states presented in the previous paragraph, a natural way to determine the amount of squeezing in Eq.~\eqref{Eq:single:mode:approx} is by identifying a direction in phase space along which there is a decrease in uncertainty in the corresponding photonic quadrature. To achieve this, we define a phase-dependent quadrature $\hat{X}_q(\theta) = \hat{X}_{q} \cos(\theta) + \hat{\bar{X}}_q\sin(\theta)$, with $\hat{X}_q = (\hat{a}_q + \hat{a}^\dagger_q)/\sqrt{2}$ and $\hat{\bar{X}}_q = (\hat{a}_q^\dagger - \hat{a}_q)/(i\sqrt{2})$, and define the optimal squeezing direction $\theta^*$ as that satisfying
	\begin{equation}\label{Eq:minimization:single}
		\Delta X_q(\theta^*)
		= \min_{\theta}[\Delta X_q(\theta)],
	\end{equation}
	where the expected values involved in the variance are taken with respect to the state in Eq.~\eqref{Eq:single:mode:approx}.
	
	The results from the optimization in Eq.~\eqref{Eq:minimization:single} are depicted in Fig.~\ref{Fig:Amount:Squeezing} for harmonic modes smaller than $q=5$. Beyond this value, the obtained results were below the numerical precision of the used software.~In panel (a) we display the optimal values of the quadrature uncertainties (blue and orange curves) and their product (black dashed line), which is expected to remain constant everywhere and equal to $1/4$.~However, the uncertainties along the different quadratures do not uniformly reach $1/2$ for all harmonic modes (gray dotted line), exhibiting peaks for harmonics $q=1$ (fundamental) and $q=3$.~This suggests the presence of significant squeezing for these specific harmonic modes.~This observation is reinforced in panel (b), where the squeezing parameter $r = -(1/2) \log_{10}(2 \Delta X_q(\theta^*))$ is plotted for the different harmonic modes. A smaller value of $r$ indicates less pronounced squeezing features in our state.~These plots reveal additional peaks for even harmonic orders, albeit diminishing as $q$ increases.
	
	\begin{figure}
		\centering
		\includegraphics[width=1\columnwidth]{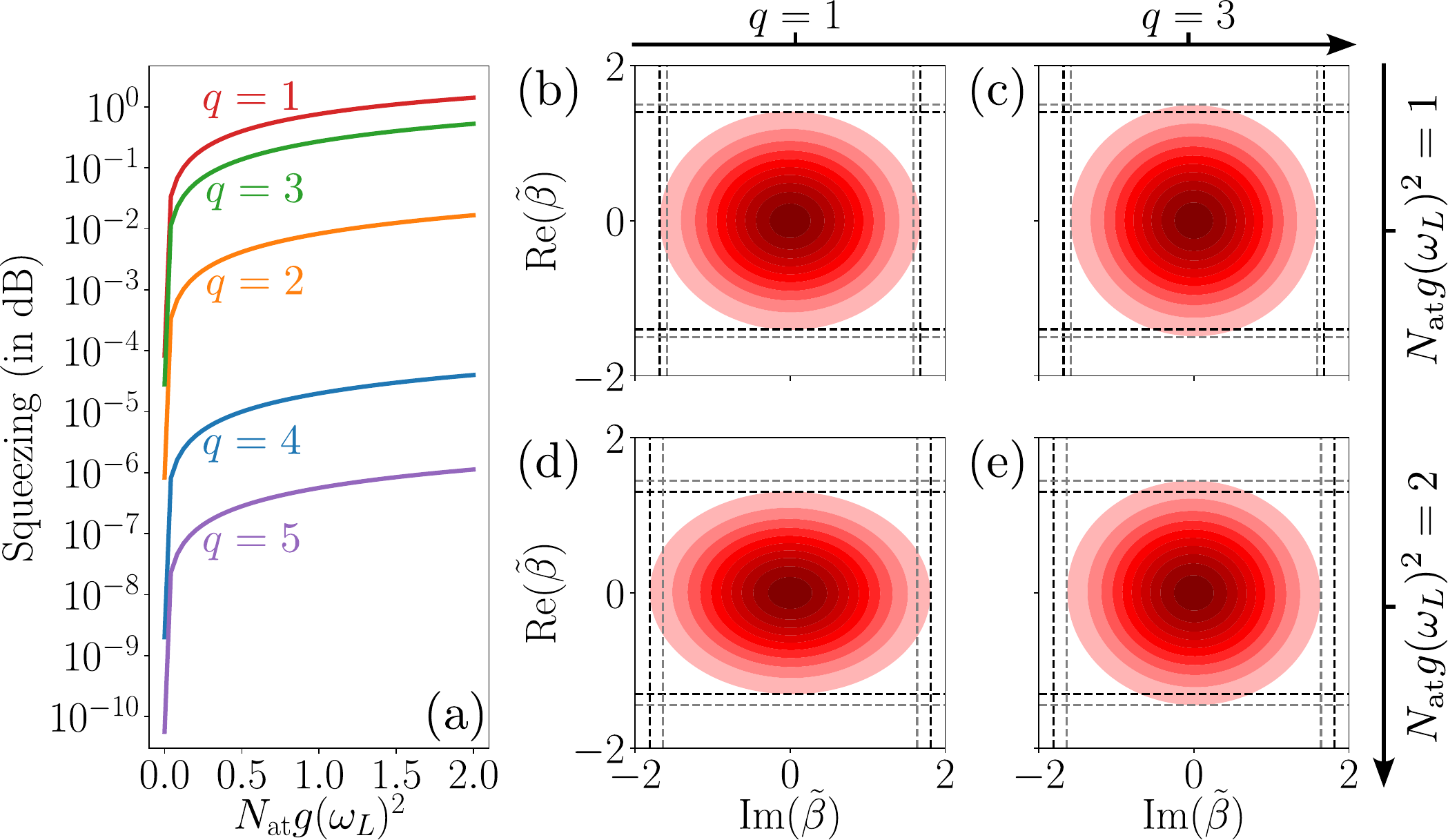}
		\caption{In panel (a), the amount of squeezing as a function of $N_{\text{at}}g(\omega_L)^2$ for different harmonic modes.~The results are presented in dB, related to the squeezing parameter $r$ through  $[\text{dB}] = 10\log_{10}(e^{2\abs{r}})$, and for values of $N_{\text{at}}g(\omega_L)^2 \in [10^{-4}, 2]$.~In panel (b), the Wigner functions of the obtained states are displayed for various values of $q$ and $N_{\text{at}}g(\omega_L)^2$, the latter represented in atomic units.~The black and grey dashed lines compare the squeezing between the leftmost and rightmost Wigner functions, respectively.}
		\label{Fig:Wigners:Squeezing}
	\end{figure}
	
	The calculations leading to Fig.~\ref{Fig:Amount:Squeezing} were performed with $N_{\text{at}}g(\omega_L)^2 = 1$. However, it is anticipated that increasing this quantity would enhance the squeezing effects.~This enhancement is indeed illustrated in Fig.~\ref{Fig:Wigners:Squeezing}, where panel (a) presents the amount of squeezing (in dB units) as a function of $N_{\text{at}}g(\omega_L)^2$.~The plot shows an increase in the total amount of squeezing for larger values of this parameter, as further exemplified in the Wigner function representations for modes $q=1$ (panels (b) and (d) with squeezing of 0.763 and 1.411 dB respectively) and $q=3$ (panels (c) and (e) with squeezing of 0.277 and 0.526 dB respectively).~In these subplots, the vertical and horizontal dashed lines delineate the quadrature uncertainties for modes $q=1$ (black lines) and $q=3$ (gray lines) for each value of $N_{\text{at}}g(\omega_L)^2$.~It is important to note that excessively large values of this parameter would necessitate consideration of the influence of the second term in Eq.~\eqref{Eq:Sol:QO:e:1}.~Additionally, $N_{\text{at}}$ represents the number of atoms initially in the excited state, and its value depends on the initial excitation conditions, including the fidelity of the employed $\pi$-pulse, and its time delay relative to the subsequent strong-laser field.~For example, if this delay surpasses the excited state lifetime, the squeezing features will be absent.

	\subsection{Two-mode squeezing}
	When referring to two-mode squeezed states, we are describing states that exhibit correlations between two different modes, denoted as $q_1$ and $q_2$, resulting in uncertainties along certain components for each mode that are reduced below those of the vacuum state. Thus, in this scenario, we extend Eq.~\eqref{Eq:single:mode:approx} to include correlations between modes $q_1$ and $q_2$
	\begin{equation}\label{Eq:two:mode:approx}
		\lvert \bar{\Phi}_{\text{e}}(t)\rangle
		\approx \hat{\vb{D}}\big(\bar{\boldsymbol{\chi}}\big)
		\exp[-\dfrac{iN_{\text{at}}}{\hbar}
		\int^t_{t_0}
		\dd \tau \hat{Q}_{\{q_1,q_2\}}(\tau)] \ket{\bar{0}},
	\end{equation}
	where $\hat{Q}_{\{q_1,q_2\}} \equiv \sum_{i,j=1}^2\hat{Q}_{q_i,q_j}$.~In this expression, we are only considering correlations between modes $q_1$ and $q_2$, while neglecting the influence of all others.~It will be shown later in this section that, similar to the amount of single-mode squeezing, the presence of correlations becomes important only for harmonics $q=1$ and $3$.
	
	One of the main advantages of two-mode squeezed states, similarly to single-mode squeezed states, is that their properties can be entirely characterized through the covariance matrix $\sigma_{q_1,q_2}$~\cite{weedbrook_gaussian_2012}, which is defined as
	\begin{equation}\label{Eq:Cov:matrix}
		\sigma_{q_1,q_2} = \mqty( A & C \\ C^T & B),
	\end{equation}
	where $A$, $B$ and $C$ are $2\times 2$ matrices whose elements are given by
	\begin{align}
		& A_{i,j} 
		= \dfrac{\langle \hat{X}_{i,q_1} \hat{X}_{j,q_1}\rangle + \langle \hat{X}_{j,q_1} \hat{X}_{i,q_1}\rangle}{2}
		- \langle \hat{X}_{i,q_1}\rangle \langle \hat{X}_{j,q_1}\rangle,
		\\
		& B_{i,j} 
		= \dfrac{\langle \hat{X}_{i,q_2} \hat{X}_{j,q_2}\rangle + \langle \hat{X}_{j,q_2} \hat{X}_{i,q_2}\rangle}{2}
		- \langle \hat{X}_{i,q_2}\rangle \langle \hat{X}_{j,q_2}\rangle,
		\\
		& C_{i,j} 
		= \langle \hat{X}_{i,q_1} \hat{X}_{j,q_2}\rangle
		- \langle \hat{X}_{i,q_1}\rangle \langle \hat{X}_{j,q_2}\rangle,
	\end{align}
	with the expectation value taken with respect to the state in Eq.~\eqref{Eq:two:mode:approx}.~Here, we denote $\hat{X}_{1,q} \equiv \hat{X}_{q}$ and $\hat{X}_{2,q}\equiv  \hat{\bar{X}}_{q}$.

	\begin{figure}
		\centering
		\includegraphics[width=1\columnwidth]{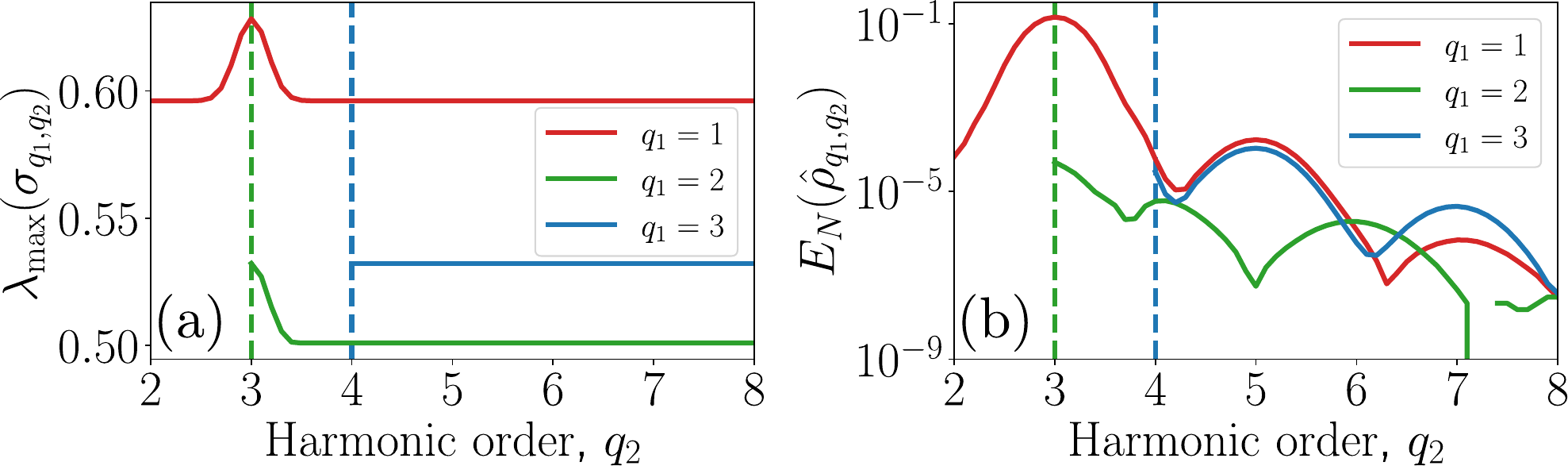}
		\caption{In panel (a), we present the maximum eigenvalue of the covariance matrix $\sigma_{q_1,q_2}$ for different pairs $(q_1,q_2)$.~The absence of squeezing features corresponds to $\lambda_{\text{max}}(\sigma) = 0.5$. In panel (b), we show the logarithmic negativity computed as in Eq.~\eqref{Eq:log:neg} for various pairs $(q_1,q_2)$.~We set $N_{\text{at}}g(\omega_L)^2 = 1$ a.u. in both plots.}
		\label{Fig:Eig:Covar}
	\end{figure}
	
	By utilizing the covariance matrix, we can equivalently perform an optimization similar to the one in Eq.~\eqref{Eq:minimization:single} by diagonalization instead of using parametrization methods.~This diagonalization enables us to identify the principal axes for the squeezing in phase space, where the different eigenvalues determine the extent of squeezing along these axes. If the maximum eigenvalue $\lambda_{\text{max}}$ equals 0.5, it indicates the absence of squeezing, while $\lambda_{\text{max}} > 0.5$ indicates squeezing is present. In Fig.~\ref{Fig:Eig:Covar}~(a), we display the maximum eigenvalue for different values of $q_1$ and $q_2$.~Despite the presence of peaks at $q_2 = 3$, the maximum eigenvalue remains constant and varies depending on $q_1$.~This reflects the single-mode squeezing trends shown in Fig.~\ref{Fig:Amount:Squeezing}.~However, the introduction of correlations with other modes extends the total squeezing beyond this constant value.~As observed, the greatest squeezing is found between modes $q=1$ and $q=3$, although significant values can also occur between modes $q=2$ and $q=3$.
	
	Although observing peaks in $\lambda_{\text{max}}$ suggests the presence of correlations between the different modes, it does not directly quantify them.~Similarly, the phase space Wigner function representations used in the single-mode analysis (Fig.~\ref{Fig:Wigners:Squeezing}~(b)) offer limited insight into the presence of two-mode squeezing beyond revealing slightly more squeezed functions, as indicated by $\lambda_{\text{max}}$, in comparison to single-mode squeezing (see Fig.~\ref{Fig:Wigners:prior} in Appendix~\ref{App:TWS}).~As two-mode squeezing inherently involves quantum correlations between different modes, entanglement measures are essential for accurate identification~\cite{horodecki_quantum_2009}.~An example of such a measure is the Positive Partial Transpose (PPT) criterion, also known as Peres-Horodecki criterion~\cite{peres_separability_1996,horodecki_separability_1996}. Interestingly, for bipartite Gaussian states as those considered here, this criterion is a necessary and sufficient condition for separability~\cite{simon_peres-horodecki_2000}.~Therefore, based on this criterion, the logarithmic negativity~\cite{horodecki_quantum_2009} serves as an ideal entanglement measure for quantifying the correlations between modes $q_1$ and $q_2$ arising from Eq.~\eqref{Eq:two:mode:approx}.~This quantity can be computed as~\cite{horodecki_quantum_2009}
	\begin{equation}\label{Eq:log:neg}
		E_N(\hat{\rho})
		= \max\big\{0, -\log_2(2 \Tilde{\nu}_-)\big\},
	\end{equation}
	where $\Tilde{\nu}_-$ is the smallest of the two symplectic eigenvalues of the partial transpose of our covariance matrix~\cite{adesso_gaussian_2005} (see Appendix~\ref{App:TWS}). The results from this analysis are presented in Fig.~\ref{Fig:Eig:Covar}~(b) where, consistent with our previous observations, the modes most correlated are $q = 1$ and $3$. Interestingly, for odd values of $q_1$, the logarithmic negativity exhibits peaks for odd values of $q_2$ and troughs for the even ones. Conversely, the opposite trend is observed when $q_1$ is even. 
	
	\begin{figure}
		\centering
		\includegraphics[width=1\columnwidth]{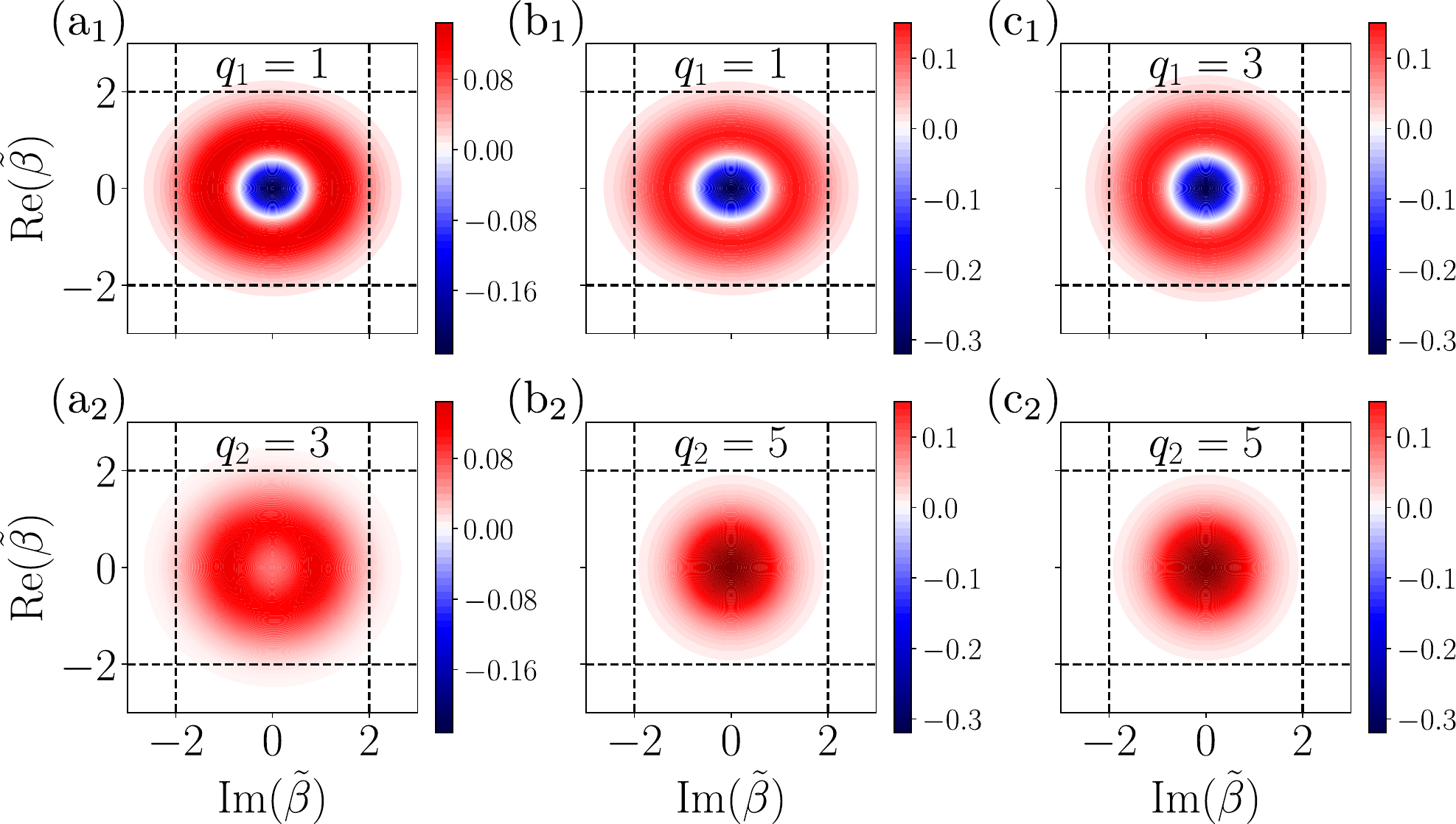}
		\caption{Wigner functions for the two-mode squeezing involving modes $q_1$ and $q_2$ after performing heralding operations in mode $q_2$ (first row) and in mode $q_1$ (second row).~For each row, the probabilities of successful heralding are as follows: in (a$_1$) $P_{\checkmark}(q_1) = 2.87 \times 10^{-3}$ and in (a$_2$) $P_{\checkmark}(q_2) = 6.09 \times 10^{-3}$; in (b$_1$) $P_{\checkmark}(q_1) = 3.51 \times 10^{-9}$ and in (b$_2$)$P_{\checkmark}(q_2) = 3.85 \times 10^{-3}$; in (c$_1$) $P_{\checkmark}(q_1) = 1.38 \times 10^{-9}$ and in (c$_2$) $P_{\checkmark}(q_2) = 4.87 \times 10^{-4}$. We set $N_{\text{at}}g(\omega_L)^2 = 1$ a.u. in all plots.}
		\label{Fig:Wigners:after}
	\end{figure}
	
	The presence of these correlations can be utilized for various purposes.~For instance, in Ref.~\cite{ourjoumtsev_generating_2006}, it was observed that performing heralding measurements on one of the involved modes allows for the generation of optical Schrödinger kitten states, i.e., superpositions of two different coherent states $\ket{\alpha}$ and $\ket{\beta}$ with $\abs{\alpha - \beta} < 2$.~This operation, termed photon subtraction, was implemented using a single-mode squeezed state, a beam splitter and photodetector heralding the presence of at least one photon in one of the output modes of the beam splitter.~In this context, the beam splitter's role was to introduce correlations between the output beams.
	
	In our case, these correlations are already present between the different modes as we are dealing with two-mode squeezed states.~Consequently, heralding operations can be implemented by physically distinguishing between different modes using a grating system, for example, and detecting the presence of at least one photon in either mode $q_1$ or $q_2$.~Mathematically, such a measurement can be represented by the projective operator $\hat{\Pi}_{q_2} = \mathbbm{1} \otimes (\mathbbm{1}-\dyad{0})$ when aiming to obtain the non-classical state in mode $q_1$, and as $\hat{\Pi}_{q_1} = (\mathbbm{1}-\dyad{0})\otimes \mathbbm{1}$ when targeting mode $q_2$. It is worth noting these operators are defined in the displaced quantum optical frame of reference $\hat{\vb{D}}(\bar{\boldsymbol{\chi}})$, and thus would require the implementation of displacement operations acting on the heralding mode~\cite{paris_displacement_1996}.~Therefore, the heralded state reads, upon normalization, as $\hat{\rho}_{q_1} = \text{tr}_{q_2}(\hat{\Pi}_{q_2} \lvert\bar{\Phi}_{\text{e}}(t)\rangle\!\langle\bar{\Phi}_{\text{e}}(t)\rvert)$ when performing the heralding on mode $q_2$.
	
	The results out of this process are presented in Fig.~\ref{Fig:Wigners:after} for different combinations of $q_1$ and $q_2$, with the success probability of heralding $P_{\checkmark}(q_1) = \text{tr}(\hat{\Pi}_{q_2}\lvert\bar{\Phi}_{\text{e}}(t)\rangle\!\langle\bar{\Phi}_{\text{e}}(t)\rvert)$ shown in the caption.~As observed, this approach leads to non-classical Wigner functions with negative regions, especially pronounced in the first row, whereas in the second row (panel (a$_2$)), the non-classical features are less pronounced, and even vanish in some cases (panels (b$_2$) and (c$_2$)).~This corresponds to cases where correlations are weaker, with the state being predominantly dominated by a (displaced) vacuum component (see Appendix~\ref{App:TWS}).~As the correlations between the involved modes weaken, the success probability of heralding decreases significantly, as seen for modes $q_1 =3$ and $q_2 = 5$.
	
	It is worth noting that, despite the similarity of the optical Schrödinger kitten-like states in the first row of Fig.~\ref{Fig:Wigners:after} to those reported in Refs.~\cite{lewenstein_generation_2021,rivera-dean_strong_2022,stammer_high_2022,stammer_theory_2022,stammer_quantum_2023}, it is the heralding protocol that sets these results apart.~In the referenced studies, optical Schrödinger kitten-like features were obtained by postselecting measured data based on energy conservation relations resulting from HHG processes~\cite{tsatrafyllis_high-order_2017,moiseyev_non-hermitian_2024,rivera-dean_quantum_2024,stammer_energy_2024}.~In contrast, our approach here leverages the entanglement features inherent to two-mode squeezing, allowing a given harmonic mode to serve as herald for the non-classical state generated in the additional mode.
	
	\section{DISCUSSION}
	While this work investigates single-mode and two-mode squeezing in various harmonic modes in HHG driven atomic systems initially in their first excited state, it is essential to highlight its novel contributions relative to similar state-of-the-art studies. This comparative perspective can help clarify the unique aspects of this study and its place within the broader field.
	
	Despite the presence of two-mode squeezing features, the results presented here differ from those of Ref.~\cite{theidel_evidence_2024} both in terms of the matter system used and the underlying dynamics leading to two-mode squeezing.~In Ref.~\cite{theidel_evidence_2024}, the non-classical features seemingly arose from Bloch oscillations within a specific band of the solid-state system~\cite{gonoskov_nonclassical_2024}.~In contrast, the two-mode squeezing is attributed to the cross-talk between the atomic ground and first excited states, a mechanisms more similar to the origin of squeezing in Ref.~\cite{stammer_squeezing_2023} and of non-classical features in Refs.~\cite{yi_generation_2024,rivera-dean_quantum-optical_2024}.~Unlike Ref.~\cite{stammer_squeezing_2023}, however, this study focuses on interactions between different bound states, allowing us to observe squeezing features without significant atomic depletion due to tunneling~\cite{delone_tunneling_1998,tong_empirical_2005}. 
	
	Although our setup involves excited states, it differs from that in Ref.~\cite{yi_generation_2024} in that we initially drive the electron to the excited state before it interacts with the strong-field.~Additionally, the non-classical features observed in our work arise from second-order effects in $g(\omega_L)$, whereas the effects in Ref.~\cite{yi_generation_2024} and those in Ref.~\cite{rivera-dean_quantum-optical_2024} originate from first-order effects. More specifically concerning Ref.~\cite{yi_generation_2024}, the observed effects depend on back-action with a resonant harmonic mode, facilitated by enclosing the interacting atoms within a cavity whose length is adjusted to achieve this resonance.
	
	Despite these distinctions, we believe that a synergistic approach combining the setup from Ref.~\cite{yi_generation_2024} with the one presented in this work would be highly interesting.~Incorporating a cavity could significantly enhance the coupling between light and matter, thereby strengthening the non-classical features. This enhancement is crucial for observing the squeezing features reported in our analysis, especially considering the condition $N_{\text{at}} g(\omega_L)^2 = 1$.~Compared to the experimental observations in Ref.~\cite{lewenstein_generation_2021}, this condition implies $N_{\text{at}}\sim 10^{16}$, while the aforementioned observations suggest contributions from $N_{\text{at}}\sim10^{12}-10^{13}$ phase-matched atoms.~However, utilizing QED cavities as in Ref.~\cite{yi_generation_2024} allows us to enhance the light-matter coupling factor $g(\omega)$ by several orders of magnitude, effectively making the generation of non-classical states of light more efficient. Alternatively, one could increase the number of atoms in the interaction region by employing high-pressure cells or optimizing the laser spot size.~Additionally, considering intermediate excited states, rather than only the ground and first excited states, could lead to larger time-dependent dipole moments, as demonstrated in Ref.~\cite{paul_enhanced_2005} to observe extended cutoffs initially reported in Refs.~\cite{watson_harmonic_1996,sanpera_harmonic-generation_1996}, which utilized a configuration similar to that of this work.

	\section{CONCLUSIONS}
	In this work, we explored a pathway for generating squeezed states of light using HHG by driving atomic systems initially prepared in their (non-degenerate) first excited state.~We characterized the amount of squeezing in both the driving field and harmonic modes, finding significant values for the low-harmonic orders and negligible amounts for the higher orders.~We also investigated two-mode squeezing features and used the generated correlations to propose heralding measurements that facilitate the generation of optical Schrödinger kitten states.
	
	This work represents an alternative pathway to existing methods for generating non-classical states of light from strong-field processes~\cite{bhattacharya_stronglaserfield_2023}.~This emerging and promising direction not only helps in further delineating the \emph{quantumness} of attosecond science~\cite{cruz-rodriguez_quantum_2024},~but also holds potential to provide unprecedented tools for generating high-intensity non-classical states of light.~These states can drive nonlinear processes in matter~\cite{lamprou_nonlinear_2023} and advance the integration of attosecond science with photonic-based quantum information science applications~\cite{lewenstein_attosecond_2024,cruz-rodriguez_quantum_2024}.
	
	\subsection*{Acknowledgements}
	ICFO group acknowledges support from: European Research Council AdG NOQIA; MCIN/AEI (PGC2018-0910.13039/501100011033, CEX2019-000910-S/10.13039/501100011033, Plan National~FIDEUA PID2019-106901GB-I00, Plan National STAMEENA PID2022-139099NB, I00, project funded by MCIN/AEI/10.13039/501100011033 and by the ``European Union NextGenerationEU/PRTR'' (PRTR-C17.I1), FPI); QUANTERA MAQS PCI2019-111828-2;  QUANTERA DYNAMITE PCI2022-132919, QuantERA II Programme co-funded by European Union’s Horizon 2020 program under Grant Agreement No 101017733; Ministry for Digital Transformation and of Civil Service of the Spanish Government through the QUANTUM ENIA project call - Quantum Spain project, and by the European Union through the Recovery, Transformation and Resilience Plan - NextGenerationEU within the framework of the Digital Spain 2026 Agenda; Fundació Cellex; Fundació Mir-Puig; Generalitat de Catalunya (European Social Fund FEDER and CERCA program, AGAUR Grant No. 2021 SGR 01452, QuantumCAT \ U16-011424, co-funded by ERDF Operational Program of Catalonia 2014-2020); Barcelona Supercomputing Center MareNostrum (FI-2023-3-0024); Funded by the European Union.~Views and opinions expressed are however those of the author(s) only and do not necessarily reflect those of the European Union, European Commission, European Climate, Infrastructure and Environment Executive Agency (CINEA), or any other granting authority.~Neither the European Union nor any granting authority can be held responsible for them (HORIZON-CL4-2022-QUANTUM-02-SGA  PASQuanS2.1, 101113690, EU Horizon 2020 FET-OPEN OPTOlogic, Grant No 899794),  EU Horizon Europe Program (This project has received funding from the European Union’s Horizon Europe research and innovation program under grant agreement No 101080086 NeQSTGrant Agreement 101080086 — NeQST); ICFO Internal ``QuantumGaudi'' project; European Union’s Horizon 2020 program under the Marie Sklodowska-Curie grant agreement No 847648;  the European Union's HORIZON MSCA-2023-DN-01 project QU-ATTO under the Marie Skłodowska-Curie Grant Agreement No. 101168628.
	
	M.~K.~and H.~B.~C.~thank the Helen Diller Quantum Center for partial financial support.
	
	P. Stammer acknowledges support from: The European Union’s Horizon 2020 research and innovation programme under the Marie Skłodowska-Curie grant agreement No 847517.
	
	P.T.'s group at FORTH acknowledges support from the Hellenic Foundation for Research and Innovation (HFRI) and the General Secretariat for Research and Technology (GSRT) under Grant Agreement CO2toO2 No. 015922, LASERLABEUROPE V (H2020-EU.1.4.1.2 Grant No. 871124), the European Union's HORIZON MSCA-2023-DN-01 project QU-ATTO under the Marie Skłodowska-Curie Grant Agreement No. 101168628, the LASERLABEUROPE V (H2020-EU.1.4.1.2 Grant No. 871124).
	
	E.P. acknowledges Royal Society fellowship funding under URF$\backslash$R1$\backslash$211390.
	
	M.~F.~C.~acknowledges support from: the National Key Research and Development Program of China (Grant No. 2023YFA1407100), the Guangdong Province
	Science and Technology Major Project (Future functional materials under extreme conditions - 2021B0301030005) and the Guangdong Natural Science Foundation (General Program project No. 2023A1515010871).
	
	\bibliography{References.bib}{}
	
	\clearpage
	\onecolumngrid
	\appendix
	
	\begin{center}
		\large{\textbf{\textsc{Appendix}}}
	\end{center}
	\section{Describing the light-matter interaction from a single-atom perspective}\label{App:Derivation}
	We begin this analysis by writing the time-dependent Schrödinger equation describing the interaction of an atomic system with a quantized electromagnetic field.~This is done within the length gauge, and working under the single-active-electron and dipole approximations.~In the interaction picture with respect to the semiclassical light-matter interaction Hamiltonian $H_{\text{sc}}(t) = \hat{H}_{\text{at}} + E_{\text{cl}}(t) \hat{d}$ (see, e.g., Refs.~\cite{rivera-dean_strong_2022,stammer_quantum_2023} for details), this equation can be written as
	\begin{equation}\label{Eq:App:Schr:Eq}
		i \hbar \pdv{\ket{\Psi(t)}}{t}
		= \hat{E}(t)\cdot \hat{d}(t) \ket{\Psi(t)},
	\end{equation}
	with $\hat{d}(t)$ the dipole moment operator in the corresponding interaction picture, and $\hat{E}(t)$ the time-dependent electric field operator, given by
	\begin{equation}\label{Eq:App:Field:Op}
		\hat{E}(t) 
		= \sum_{q=1}^{q_c} \hat{E}_{q}(t)
		= -i f(t) \sum_{q=1}^{q_c}
		g(\omega_q)
		\big[
		\hat{a}e^{-i\omega_q t}
		- \hat{a}^\dagger e^{i\omega_q t}
		\big],
	\end{equation}
	where we discretize the electromagnetic field modes and account for the pulsed nature of the employed source by means of an envelope function $0 \leq f(t) \leq 1$. While a fully rigorous quantum optical analysis would typically require an infinite number of modes to represent a continuous spectrum, we apply the envelope function $f(t)$ both to confine the interaction to a finite time interval and to recover the classical-pulsed field expressions. This approach yields results equivalent to those obtained when using multimode descriptions of the driving laser field~\cite{rivera-dean_strong_2022,stammer_quantum_2023}, while preserving the simplicity of using single-mode drivers.
	
	The main difference between this work and Refs.~\cite{rivera-dean_strong_2022,stammer_quantum_2023,stammer_squeezing_2023} is that we consider the initial state of the system to be in a superposition of the form
	\begin{equation}\label{Eq:App:initial:state}
		\ket{\Psi(t_0)}
		= \big[
		c_{\text{g}}
		\ket{\text{g}}
		+ c_{\text{e}}
		\ket{\text{e}}
		\big]
		\bigotimes_{q=1}^{q_c} \ket{0_q},
	\end{equation}
	that is, the atomic system is in a superposition of different energetic states, where $c_{\text{g}}$ and $c_{\text{e}}$ represent the initial probability amplitudes associated with the ground and first excited states, respectively.~Here, we consider the initial electronic state to be an arbitrary superposition of the ground and excited states, unlike in the main text. At the end of this analysis, we impose the initial condition$c_g=0$ and $c_e = 1$, which yield Eqs.~\eqref{Eq:Solution:ground} and \eqref{Eq:Solution:excited} in the main text.
	
	To solve the differential equation above, we introduce the identity in the electronic subspace as 
	\begin{equation}\label{Eq:App:identity}
		\mathbbm{1}
		= \dyad{\text{g}}
		+ \dyad{\text{e}}
		+ \sum_{n=2}\dyad{\psi_n}
		+ \int \dd \psi_c \dyad{\psi_c},
	\end{equation}
	where the first two terms are projectors with respect to the ground and first excited states of the atomic system, respectively, which we assume non-degenerate.~The third term includes the projector onto all other bound states, and the last term accounts for all continuum states. In the following analysis, we neglect the contribution of continuum states, as their impact is small compared to that of the atomic lowest energy states for the field parameters we consider~\cite{stammer_squeezing_2023}. However, it is important to highlight that these continuum states have indeed been taken into account to compute the time-dependent dipole moment matrix elements, as these are evaluated in the original electronic frame of reference.~Furthermore, contrary to the standard Strong-Field Approximation (SFA) assumptions that disregard contributions from all bound states other than the ground state~\cite{lewenstein_theory_1994}, we include the contribution of the first excited state.~This inclusion is crucial, as the initial state of the atom, as described in Eq.~\eqref{Eq:App:initial:state}, makes the contributions from excited states significant in defining the properties of the generated harmonic radiation~\cite{watson_harmonic_1996,sanpera_harmonic-generation_1996}.
	
	\begin{figure}[b]
		\centering
		\includegraphics[width=0.8\textwidth]{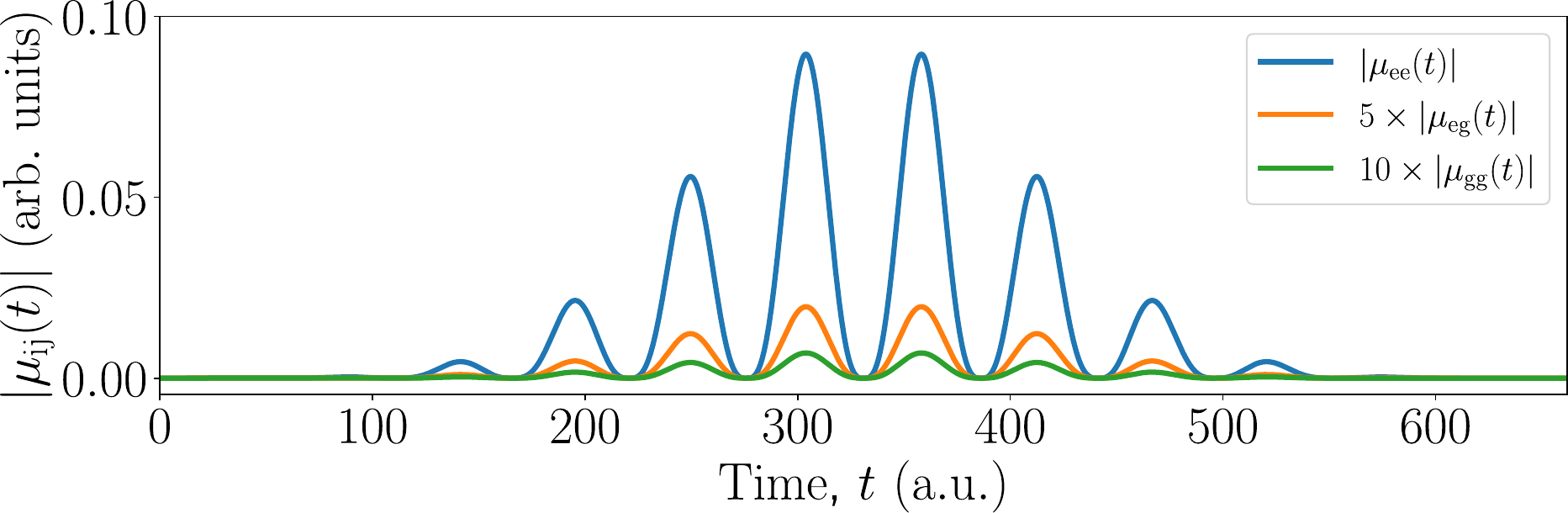}
		\caption{Behavior of the matrix elements of the time-dependent dipoles as a function of time. Here, we considered a field of amplitude $E_0 = 0.053$ a.u., $\omega_L = 0.057$ a.u., with a $\sin^2$-envelope and duration of 6 optical cycles. The target atom used in the analysis is He$^+$, which has ionization potentials $I_{p,\text{g}} = 2$ a.u. and $I_{p,\text{e}} = 0.5$ a.u.  for the ground and first excited states, respectively. In the plots, some of the matrix elements have been multiplied by a constant to properly observe their behavior.}
		\label{Fig:dipoles}
	\end{figure}
	
	Therefore, introducing Eq.~\eqref{Eq:App:identity} into the Schrödinger equation \eqref{Eq:App:Schr:Eq} and taking into account the simplications above, we arrive to the following set of coupled differential equations
	\begin{align}
		&i\hbar \pdv{\ket{\Phi_{\text{g}}(t)}}{t}
		= \mu_{\text{gg}}(t)\hat{E}(t)
		\ket{\Phi_{\text{g}}(t)}
		+ \mu_{\text{ge}}(t)\hat{E}(t)
		\ket{\Phi_{\text{e}}(t)},\label{Eq:App:TDSE:g}
		\\&
		i\hbar \pdv{\ket{\Phi_{\text{e}}(t)}}{t}
		= \mu_{\text{eg}}(t)\hat{E}(t)
		\ket{\Phi_{\text{g}}(t)}
		+ \mu_{\text{ee}}(t)\hat{E}(t)
		\ket{\Phi_{\text{e}}(t)},\label{Eq:App:TDSE:e}
	\end{align}
	where $\ket{\Phi_{\text{i}}(t)} \equiv \braket{\text{i}}{\Psi(t)}$ and $\mu_{\text{ij}}(t) \equiv \bra{\text{i}}\hat{U}^\dagger_{\text{sc}}(t,t_0) \hat{r}\hat{U}_{\text{sc}}(t,t_0)\ket{\text{j}}$ are the time-dependent dipole moment matrix elements. These matrix elements are computed following Ref.~\cite{watson_harmonic_1996,sanpera_harmonic-generation_1996}, by combining numerical integration with saddle-point approaches. More specifically, 3D integrals involving momentum variables have been evaluated using the saddle-point approximation, while the time-dependent integrals numerically using adaptive algorithms.~Different laser parameters were used compared to the aforementioned references to make the forthcoming numerical calculations less time-consuming.~The absolute values of these matrix elements are presented in Fig.~\ref{Fig:dipoles}, where it is observed that $\abs{\mu_{\text{ee}}(t)} \geq \abs{\mu_{\text{eg}}} \geq \abs{\mu_{\text{gg}}(t)}$, in agreement with those of Ref.~\cite{gorlach_quantum-optical_2020}.
	
	Given that $\mu_{\text{gg}}(t)$ is perturbatively smaller compared to the other two, we apply perturbation theory to $\ket{\Phi_{\text{g}}(t)}$
	\begin{equation}\label{Eq:App:gg:perb}
		\ket{\Phi_{\text{g}}(t)}
		= \lvert\Phi^{(0)}_{\text{g}}(t)\rangle
		+ \sum_{n=1}^{\infty} \lvert\Phi^{(n)}_{\text{g}}(t)\rangle,
	\end{equation}
	such that, when substituted in Eq.~\eqref{Eq:App:TDSE:g}, it yields the following set of coupled differential equations for each perturbative order
	\begin{align}
		&i\hbar \pdv{\lvert\Phi^{(0)}_{\text{g}}(t)\rangle}{t}
		= \mu_{\text{ge}}(t)\hat{E}(t)
		\ket{\Phi_{\text{e}}(t)},
		\\& i\hbar \pdv{\lvert\Phi^{(n)}_{\text{g}}(t)\rangle}{t}
		= \mu_{\text{gg}}(t)\hat{E}(t)
		\lvert\Phi^{(n-1)}_{\text{g}}(t)\rangle  \quad \forall n> 0,
	\end{align}
	whose solutions are straightforwardly given by
	\begin{align}
		& \lvert\Phi^{(0)}_{\text{g}}(t)\rangle
		=\lvert\Phi^{(0)}_{\text{g}}(t_0)\rangle
		-\dfrac{i}{\hbar}
		\int^{t}_{t_0} \dd \tau
		\mu_{\text{ge}}(\tau) \hat{E}(\tau)
		\ket{\Phi_{\text{e}}(\tau)}, \label{Eq:App:ground:zeroth}
		\\&	\lvert\Phi^{(n)}_{\text{g}}(t)\rangle
		=\lvert\Phi^{(n)}_{\text{g}}(t_0)\rangle
		-\dfrac{i}{\hbar}
		\int^{t}_{t_0} \dd \tau
		\mu_{\text{gg}}(\tau) \hat{E}(\tau)
		\lvert\Phi^{(n-1)}_{\text{g}}(t)\rangle \quad \forall n> 0.
	\end{align}
	
	We now substitute this solution into Eq.~\eqref{Eq:App:TDSE:e}. Given that $\abs{\mu_{\text{eg}}(t)}$ can be treated as a perturbative quantity with respect to $\abs{\mu_{\text{ee}}(t)}$, we retain only the zeroth-order solution of Eq.~\eqref{Eq:App:gg:perb}.~Consequently, the resulting differential equation can be approximately expressed as
	\begin{equation}
		i \hbar \pdv{\ket{\Phi_{\text{e}}(t)}}{t}
		\approx \mu_{\text{ee}}(t) \hat{E}(t) \ket{\Phi_{\text{e}}(t)}
		+ \mu_{\text{eg}}(t)\hat{E}(t)
		\lvert\Phi^{(0)}_{\text{g}}(t_0)\rangle
		- \dfrac{i}{\hbar}\mu_{\text{eg}}(t)\hat{E}(t)
		\int^{t}_{t_0} \dd \tau
		\mu_{\text{ge}}(\tau) \hat{E}(\tau)
		\ket{\Phi_{\text{e}}(\tau)},
	\end{equation}
	and, similarly to Ref.~\cite{stammer_squeezing_2023}, we apply the Markov approximation, which allows us to express the equation above as
	\begin{equation}
		i \hbar \pdv{\ket{\Phi_{\text{e}}(t)}}{t}
		\approx 
		\bigg[
		\mu_{\text{ee}}(t) \hat{E}(t)
		- \dfrac{i}{\hbar}\mu_{\text{eg}}(t)\hat{E}(t)
		\int^{t}_{t_0} \dd \tau
		\mu_{\text{ge}}(\tau) \hat{E}(\tau)
		\bigg]\ket{\Phi_{\text{e}}(t)}
		+  \mu_{\text{eg}}(t)\hat{E}(t)
		\lvert\Phi^{(0)}_{\text{g}}(t_0)\rangle,
	\end{equation}
	which is an inhomogeneous first-order differential equation with distinct homogeneous and inhomogeneous parts. A solution to this equation can be expressed as the sum of a solution to the homogeneous part and a particular solution to the inhomogeneous part.~Therefore, we first focus on the homogeneous differential equation, which is given by
	\begin{equation}
		i \hbar \pdv{\lvert\Phi^{(\text{h})}_{\text{e}}(t)\rangle}{t}
		\approx 
		\bigg[
		\mu_{\text{ee}}(t) \hat{E}(t)
		- \dfrac{i}{\hbar}\mu_{\text{eg}}(t)\hat{E}(t)
		\int^{t}_{t_0} \dd \tau
		\mu_{\text{ge}}(\tau) \hat{E}(\tau)
		\bigg
		]\lvert\Phi^{(\text{h})}_{\text{e}}(t)\rangle,
	\end{equation}
	and whose solution can be generally written, for small enough values of $\Delta t$, as
	\begin{equation}\label{Eq:App:Trotter}
		\lvert\Phi^{(\text{h})}_{\text{e}}(t)\rangle
		= \lim_{N\to\infty}
		\Bigg\{
		\prod^N_{n=1}
		\exp[-\dfrac{i}{\hbar} \hat{O}(t_n)\Delta t] 
		\Bigg\}
		\lvert\Phi^{(\text{h})}_{\text{e}}(t_0)\rangle,
	\end{equation}
	where we define
	$
	\hat{O}(t)
	\equiv \mu_{\text{ee}}(t) \hat{E}(t)
	- \hat{Q}(t)$ and 
	$
	\hat{Q}(t)
	\equiv 
	\frac{i}{\hbar}\mu_{\text{eg}}(t)\hat{E}(t)
	\int^{t}_{t_0} \dd \tau
	\mu_{\text{ge}}(\tau) \hat{E}(\tau)
	$. Our main objective is to express the exponential operator as the product of two exponential operators, each incorporating the effects of $\mu_{\text{ee}}(t) \hat{E}(t)$ and $\hat{Q}(t)$, respectively.~This involves using the Baker-Campbell-Hausdorff (BCH) formula.~To achieve this, we first evaluate the commutation between these operators at different times.
	
	We begin by evaluating the commutator $[\hat{E}(t),\hat{E}(t')]$. Considering the definition of the electric field operator given in Eq.~\eqref{Eq:App:Field:Op}, and using the commutation relation $[\hat{a}_q, \hat{a}^\dagger_{q'}] = \delta_{q,q'}$, we can write
	\begin{equation}
		[\hat{E}(t),\hat{E}(t')]
		= -i2f(t)f(t')
		\sum_{q=1}^{q_c}
		\sin(\omega_q(t-t'))
		\equiv i\sum_{q=1}^{q_c} \varphi_q(t',t),
	\end{equation}
	which is essentially a complex constant that depends on $g(\omega_q)^2$ multiplied by the identity. This expression can be used to evaluate the commutator $[\hat{E}(t),\hat{Q}(t')]$, which can be expressed as
	\begin{equation}
		[\hat{E}(t),\hat{Q}(t)]
		= \dfrac{i}{\hbar} \mu_{\text{eg}}(t') 
		\int^{t'}_{t_0} \dd \tau
		\mu_{\text{ge}}(\tau)
		[\hat{E}(t), \hat{E}(t')\hat{E}(\tau)],
	\end{equation}
	with the commutator in the integrand satisfying
	\begin{equation}
		[\hat{E}(t), \hat{E}(t')\hat{E}(\tau)]
		= \hat{E}(t') [\hat{E}(t),\hat{E}(\tau)]
		+ [\hat{E}(t),\hat{E}(t')]\hat{E}(\tau)
		= i\sum_q
		\big[
		\varphi_q(t,\tau)\hat{E}(t') + \varphi_q(t,t') \hat{E}(\tau)
		\big]
		\propto \hat{\mathcal{O}}(g^3(\omega_q)).
	\end{equation}
	
	Thus, if we restrict ourselves to cases where terms of order $\mathcal{O}(g(\omega_q)^3)$ are negligible, we can approximate $[\hat{E}(t),\hat{Q}(t')] \propto \hat{\mathcal{O}}(g^3(\omega_q)) \approx 0$. Similarly, one can show that $[\hat{Q}(t),\hat{Q}(t')] \propto \hat{\mathcal{O}}(g^4(\omega_q)) \approx 0$.~Consequently, using the BCH formula and incorporating these approximations, we can express Eq.~\eqref{Eq:App:Trotter} as
	\begin{equation}
		\begin{aligned}
			\hat{U}(t,t_0)
			&\approx \lim_{N\to\infty}
			\bigg\{
			\prod_{n=1}^N 
			\exp[-\dfrac{i}{\hbar} \mu_{\text{ee}}(t_n)\Delta t]
			\exp[-\dfrac{i}{\hbar}\hat{Q}(t_n)\Delta t]
			\\&\approx e^{i\varphi(t)}
			\exp[-\dfrac{i}{\hbar} \int^{t}_{t_0} \dd \tau \mu_{\text{ee}}(\tau) \hat{E}(\tau)]
			\exp[-\dfrac{i}{\hbar} \int^{t}_{t_0} \dd \tau \hat{Q}(\tau)]
			\bigg\},
		\end{aligned}
	\end{equation}
	where we have denoted $\lvert\Phi_{\text{e}}^{(\text{h})}(t)\rangle = \hat{U}(t,t_0) \lvert\Phi_{\text{e}}^{(\text{h})}(t_0)\rangle$. With this solution, we can then write the solution to the inhomogeneous equation
	\begin{equation}
		\ket{\Phi_{\text{e}}(t)}
		= \hat{U}(t,t_0) \lvert\Phi_{\text{e}}(t_0)\rangle
		+\dfrac{i}{\hbar}
		\int^{t}_{t_0} \dd \tau
		\hat{U}(t,\tau) 
		\mu_{\text{eg}}(\tau)
		\hat{E}(\tau) \lvert\Phi^{(0)}_{\text{g}}(\tau)\rangle,
	\end{equation}
	and, similarly, we get for the zeroth order term in Eq.~\eqref{Eq:App:ground:zeroth}
	\begin{equation}
		\begin{aligned}
			\lvert\Phi^{(0)}_{\text{g}}(t)\rangle
			&=\lvert\Phi^{(0)}_{\text{g}}(t_0)\rangle
			- \dfrac{i}{\hbar}
			\int^t_{t_0} \dd \tau \mu_{\text{ge}}(\tau) \hat{E}(\tau)
			\hat{U}(\tau,t_0) \lvert\Phi_{\text{e}}(t_0)\rangle
			\\&\quad
			+ \dfrac{1}{\hbar^2}
			\int^t_{t_0} \dd \tau_1
			\mu_{\text{ge}}(\tau_1) \hat{E}(\tau_1)
			\int^{\tau_1}_{t_0} \dd \tau_2
			\hat{U}(\tau_1,\tau_2) 
			\mu_{\text{eg}}(\tau_2)
			\hat{E}(\tau_2) \lvert\Phi^{(0)}_{\text{g}}(\tau_2)\rangle
			\\ & \approx
			\lvert\Phi^{(0)}_{\text{g}}(t_0)\rangle
			- \dfrac{i}{\hbar}
			\int^t_{t_0} \dd \tau \mu_{\text{ge}}(\tau)  \hat{E}(\tau)
			\hat{U}(\tau,t_0) \lvert\Phi_{\text{e}}(t_0)\rangle
			\\&\quad
			+ \dfrac{1}{\hbar^2}
			\int^t_{t_0} \dd \tau_1
			\mu_{\text{ge}}(\tau_1) \hat{E}(\tau_1)
			\int^{\tau_1}_{t_0} \dd \tau_2
			\hat{U}(\tau_1,\tau_2) 
			\mu_{\text{eg}}(\tau_2)
			\hat{E}(\tau_2) \lvert\Phi^{(0)}_{\text{g}}(t_0)\rangle,
		\end{aligned}
	\end{equation}
	where in the last approximation we have omitted the contribution of $\mathcal{O}(g^3(\omega_q))$ terms.
	
	At this point, we can introduce the initial conditions, which in our case are $\ket{\Phi_{\text{e}}(t_0)} = c_{\text{e}}\ket{\bar{0}}$, $\lvert\Phi^{(0)}_{\text{g}}(t_0)\rangle = c_{\text{g}}\ket{\bar{0}}$ and $\lvert\Phi^{(n)}_{\text{g}}(t_0)\rangle = 0$. More specifically, in the main text we are interested in the regime where all the population is initially in the excited state, i.e. $c_{\text{e}} = 1$ and $c_{\text{g}} = 0$. From this, we can write
	\begin{align}
		&\ket{\Phi_{\text{e}}(t)}
		\approx \hat{U}(t,t_0)\ket{\bar{0}}
		+ \dfrac{1}{\hbar^2}
		\int^t_{t_0} \dd \tau_1
		\int^{\tau_1}_{t_0} \dd \tau_2
		\hat{U}(t,\tau_1) \mu_{\text{eg}}(\tau_1) \hat{E}(\tau_1)
		\mu_{\text{ge}}(\tau_2) \hat{E}(\tau_2) \hat{U}(\tau_2,t_0)\ket{\bar{0}},\label{Eq:App:Sol:e:t0}
		\\&
		\lvert \Phi^{(0)}_{\text{g}}(t)\rangle
		\approx - \dfrac{i}{\hbar}
		\int^t_{t_0} \dd \tau
		\mu_{\text{ge}}(\tau)
		\hat{E}(\tau)
		\hat{U}(\tau,t_0) \ket{\bar{0}}.
		\label{Eq:App:Sol:g:t0}
	\end{align}
	
	As seen in Eq.~\eqref{Eq:App:Sol:e:t0}, squeezing features are most pronounced when the electron ends in the (dressed) excited state.~In contrast, outcomes where the electron is found in the (dressed) ground state produce delocalized single-photon excitations across the various harmonic modes.~This quantum optical state closely resembles those in Ref.\cite{rivera-dean_quantum-optical_2024}, particularly when focusing on events where the electron ends in the antibonding (first excited) state.~This configuration shows significant entanglement and other non-classical properties, making further analysis of Eq.~\eqref{Eq:App:Sol:g:t0} redundant for this work; interested readers are directed to the reference above.
	
	Similarly, if we consider initial conditions where $c_{\text{g}} = 1$ and $c_{\text{e}} = 0$, while restricting to scenarios where the electron ultimately occupies the first excited state, we derive
	\begin{equation}
		\ket{\Phi_{\text{e}}(t)}
		\approx
		\dfrac{i}{\hbar}
		\int^{t}_{t_0} \dd \tau
		\hat{U}(t,\tau) 
		\mu_{\text{eg}}(\tau)
		\hat{E}(\tau) \lvert\bar{0}\rangle,
	\end{equation}
	which suggests the presence of squeezed single-photon excitations across harmonic modes.~However, these features are minimal in the single-atom regime and would remain so in the many-atom scenario, as can be seen by following a similar analysis to that in Ref.~\cite{rivera-dean_quantum-optical_2024}, equivalent to the one presented in the next section.
	
	\section{Laboratory frame and a many-atom considerations}\label{App:MB}
	The analysis we have presented thus far has been conducted in the semiclassical framework, specifically in the interaction picture with respect to $\hat{H}_{\text{sc}}(t)$.~Within this framework, we have derived the following solution to the Schrödinger equation
	\begin{equation}\label{Eq:App:Sol}
		\ket{\Psi(t)}
		\approx \ket{\text{e}}\otimes\ket{\Phi_{\text{e}}(t)}
		+ \ket{\text{g}}\otimes\lvert\Phi^{(0)}_{\text{g}}(t)\rangle,
	\end{equation}
	where $\ket{\Phi_{\text{e}}(t)}$ and $ \lvert\Phi^{(0)}_{\text{g}}(t)\rangle$ are given in Eqs.~\eqref{Eq:App:Sol:e:t0} and \eqref{Eq:App:Sol:g:t0}, respectively. 
	
	Given that the relation between the laboratory frame and the semiclassical picture considered here is given by $\lvert\bar{\Psi}(t)\rangle = \hat{U}_{\text{sc}}(t) \ket{\Psi(t)}$, we can express the state in the electronic laboratory frame as follows
	\begin{equation}
		\lvert\bar{\Psi}(t)\rangle
		= \big(
		\hat{U}_{\text{sc}}(t)
		\ket{\text{e}}
		\big)
		\otimes\ket{\Phi_{\text{e}}(t)}
		+ \big(
		\hat{U}_{\text{sc}}(t)
		\ket{\text{g}}
		\big)
		\otimes\ket{\Phi_{\text{g}}(t)},
	\end{equation}
	and under the strong-field approximation~\cite{watson_harmonic_1996,sanpera_harmonic-generation_1996}, we can write
	\begin{align}
		&\hat{U}_{\text{sc}}(t) \ket{\text{e}}
		= e^{-i I_{p, \text{e}} t/\hbar} \ket{\text{e}}
		+ \int \dd c \ b_c(t) \ket{\phi_c},
		\\
		&\hat{U}_{\text{sc}}(t) \ket{\text{g}}
		= e^{-i I_{p, \text{g}} t/\hbar} \ket{\text{g}},
	\end{align}
	where $\{\ket{\phi_c}\}$ denotes the set of continuum states, and $I_{p, \text{g}}$ and $I_{p, \text{e}}$ denote the ionization potential for ground and first excited states, respectively. Among all possible events, we are particularly interested in those where the electron returns to its initial state, i.e., the excited state. Consequently, considering those, we find for the quantum optical state
	\begin{equation}\label{Eq:App:Final:state:cond:e}
		\begin{aligned}
			\lvert \bar{\Phi}_{\text{e}}(t)\rangle
			= \langle \text{e} \vert \bar{\Psi}(t)\rangle
			&\approx \bigg[
			\hat{U}(t,t_0)
			+ \dfrac{1}{\hbar^2}
			\int^t_{t_0} \dd \tau_1
			\int^{\tau_1}_{t_0} \dd \tau_2
			\hat{U}(t,\tau_1) \mu_{\text{eg}}(\tau_1) \hat{E}(\tau_1)
			\mu_{\text{ge}}(\tau_2) \hat{E}(\tau_2) \hat{U}(\tau_2,t_0)
			\bigg] \ket{0}
			\\
			&\equiv \hat{\mathcal{U}}(t) \ket{0},
		\end{aligned}
	\end{equation}
	corresponding to the superposition of a displaced squeezed vacuum state, a photon-added displaced squeezed vacuum state, and a photon-subtracted displaced squeezed vacuum state.
	
	However, all the analysis presented thus far has been conducted at the single-atom level.~Consequently, given the weak coupling between light and matter, the non-classical features we could potentially observe in our state are minimal. Nevertheless, in typical HHG setups, many atoms independently couple to the same electromagnetic field. In this scenario, and within the semiclassical interaction picture, the time-dependent Hamiltonian describing the system's dynamics is given by
	\begin{equation}
		\hat{H}(t)
		= \sum_{i=1}^{N_{\text{at}}}
		\hat{E}(t) \hat{d}_i(t),
	\end{equation}
	where $\hat{d}_i(t)$ is the time-dependent dipole moment operator of the $i$th atom, with $N_{\text{at}}$ the total number of atoms. It is important to note that this Hamiltonian is not local, as all the atoms couple independently to the same electromagnetic field. This ultimately leads to coupled dynamics, and, in general, we have that
	\begin{equation}\label{Eq:App:Comm:Corr}
		\big[
		\hat{E}(t) r_i(t), \hat{E}(t')r_j(t')
		\big]
		= \hat{r}_i(t)\hat{r}_j(t) 
		\big[\hat{E}(t),\hat{E}(t')\big]
		= i \hat{r}_i(t)\hat{r}_j(t) 
		\sum_{q} \varphi_q(t) \neq 0,
	\end{equation}
	where we have taken into account that $[\hat{r}_i(t),\hat{r}_j(t)] = 0$ $\forall i \neq j$.
	
	Formally, this means that to solve the many-body problem, we must consider the contributions from all atoms. However, if we assume that, on average, the time-dependent dipole moments of different atoms are uncorrelated~\cite{sundaram_high-order_1990}, i.e. $\langle \hat{r}_i(t) \hat{r}_j(t')\rangle = 0$, the contributions in Eq.~\eqref{Eq:App:Comm:Corr} average to zero.~More specifically, considering for simplicity a single harmonic mode $q$, we obtain
	\begin{equation}\label{Eq:App:Average:Corr}
		\begin{aligned}
			\int \dd t \int \dd t'
			\big\langle
			\big[
			\hat{E}_q(t) r_i(t), \hat{E}_q(t')r_j(t')
			\big]
			\big\rangle
			&= i\int \dd t \int \dd t'
			\langle \hat{r}_i(t)\hat{r}_j(t') \rangle 
			\sin(\omega_q(t-t'))
			\\&
			= \dfrac12
			\int \dd t \int \dd t'
			\langle \hat{r}_i(t)\rangle\langle\hat{r}_j(t') \rangle 
			\Big[ 
			e^{i\omega_q (t-t')}
			- e^{-i\omega_q(t-t')}
			\Big]
			\\&
			= \dfrac12
			\Bigg[
			\bigg\lvert
			\int \dd t \langle \hat{r}_i(t)\rangle e^{i\omega_q t}
			\bigg\rvert^2
			- 	\bigg\lvert
			\int \dd t \langle \hat{r}_i(t)\rangle e^{i\omega_q t}
			\bigg\rvert^2
			\Bigg]
			= 0,
		\end{aligned}
	\end{equation}
	where we have assumed that $\langle \hat{r}_i(t)\rangle = \langle \hat{r}_j(t)\rangle$ $\forall i,j$, which is valid under the dipole approximation and when all atoms are of the same species. Thus, Eq.~\eqref{Eq:App:Average:Corr} implies that the average correlations vanish. Given that the coupling of each atom with the field is weak, it is then expected that the influence of each individual atom on the field is weak enough for the interaction to be considered independent.~Consequently, if we assume that all atoms are initially in the same state, the many-body evolution can be expressed as
	\begin{equation}
		\lvert\bar{\Psi}(t)\rangle
		= \hat{\bar{\vb{U}}}(t)
		\bigotimes_{i=1}^{N_{\text{at}}}\ket{\text{e}} \otimes \ket{\bar{0}}
		\approx
		\prod^{N_{\text{at}}}_{i=1}
		\hat{\vb{U}}(t) 
		\bigotimes_{i=1}^{N_{\text{at}}}\ket{\text{e}} \otimes \ket{\bar{0}},
	\end{equation}
	and focusing on those cases where the electron ends up in the excited state, we obtain in the laboratory frame
	\begin{equation}\label{Eq:App:Squeezed:Nat}
		\lvert \bar{\Phi}_{\text{e}}(t)\rangle
		\approx 
		\big[
		\hat{\mathcal{U}}(t)
		\big]^{N_{\text{at}}} \ket{\bar{0}}.
	\end{equation}
	
	\section{Details about the single-mode squeezing analysis}\label{App:SMS}
	Provided the state in Eq.~\eqref{Eq:App:Squeezed:Nat}, we want to characterize the amount of squeezing. In principle, such states involve correlations between all possible modes, but for the moment we are going to consider the squeezing terms affecting single modes. In other words, we will neglect, for simplicity, the two-mode squeezing contributions of $\hat{Q}$ and keep only the diagonal ones, i.e.
	\begin{equation}
		\hat{Q}_{q,q}
		= \dfrac{i}{\hbar} \mu_{\text{eg}}(t) \hat{E}_q(t) \int^t_{t_0} \dd \tau \mu_{\text{ge}}(\tau) \hat{E}_{q'}(\tau),
	\end{equation}
	such that we approximate Eq.~\eqref{Eq:App:Squeezed:Nat} as
	\begin{equation}\label{Eq:App:Squeez:Single:Mode}
		\lvert\bar{\Phi}(t)\rangle
		\approx \hat{\vb{D}}\big(\bar{\boldsymbol{\chi}}\big)
		\exp[-\sum_{q=1}^{q_c}
		\dfrac{iN_{\text{at}}}{\hbar}
		\int^t_{t_0}
		\dd \tau \hat{Q}_{q,q}(\tau)] \ket{\bar{0}},
	\end{equation}
	where we defined
	$
	\hat{\vb{D}}(
	\boldsymbol{\chi}
	\big) \equiv \prod_{q=1}^{q_c} \hat{D}_q(\chi_q)
	$ with $\hat{D}(\cdot)$ representing the displacement operator and $\hat{Q}_{q,q}(t) \equiv \frac{i}{\hbar}\mu_{\text{eg}}(t) \hat{E}_q(t) \int^t_{t_0} \dd \tau \mu_{\text{ge}}(\tau) \hat{E}_q(\tau)$, which involves second-order terms of creation and annihilation operators.~In this expression, $\bar{\boldsymbol{\chi}} \equiv f(\boldsymbol{\chi}_{\text{e}}(t))$, where $f(\cdot)$ is a function that depends on the prefactors accompanying the product of the creation and annihilation operators in $\hat{Q}_{q,q}$.~This can be observed from the fact that we can always write $[f(\hat{a}^2, \hat{a}^\dagger\hat{a}), g(\hat{a})] = h(\hat{a})$, with $f(\cdot)$, $g(\cdot)$ and $h(\cdot)$ being linear functions of the corresponding operators.~Therefore, $\exp[-f(\hat{a}^2, \hat{a}^\dagger\hat{a})]\exp[g(\hat{a})] \exp[f(\hat{a}^2, \hat{a}^\dagger\hat{a})] = \text{exp}[\Tilde{h}(\hat{a})]$, where $\Tilde{h}(\cdot)$ does not necessarily have to be equal to $h(\cdot)$, but still is a linear function of creation and annihilation operators.
	
	To quantify the amount of squeezing  present in each harmonic mode of our state, we estimate the value of the squeezing parameter $r$ by solving the following optimization problem
	\begin{equation}\label{Eq:App:optimization}
		\Delta X_q(\theta^*)
		= \min_{\theta} 
		\big[
		\Delta X_q(\theta)
		\big],
	\end{equation}
	where $\hat{X}(\theta)$ is a phase-dependent photonic quadrature defined as
	\begin{equation}
		\hat{X}_{q} (\theta)
		\equiv \hat{X}_{q} \cos(\theta) + \hat{\bar{X}}_{q} \sin(\theta),
	\end{equation}
	with $\hat{X}_{q} = (\hat{a}_q + \hat{a}_q^\dagger)/\sqrt{2}$ and $\hat{\bar{X}}_{q} = (\hat{a}^\dagger_q - \hat{a}_q^\dagger)/(i\sqrt{2})$. Thus, the optimization problem in Eq.~\eqref{Eq:App:optimization} allows us to find a direction in phase space along which the corresponding distribution gets squeezed.
	
	Such optimization has been implemented numerically in Python by combining a brute force search with built-in functions from the \texttt{Qutip} package~\cite{johansson_qutip_2012,johansson_qutip_2013}. Firstly, the prefactors to the set of operators $\{\hat{a}_q^2, \hat{a}_q\hat{a}_q^\dagger, \hat{a}_q^\dagger \hat{a}_q, \hat{a}_q^{\dagger 2}\}$ were computed through numerical integration using the \texttt{quad} function of the \texttt{Scipy} package~\cite{2020SciPy-NMeth}, with the integration parameters suitably adjusted to achieve convergence. Specifically, an upper bound of 1000 subintervals was employed in the adaptive algorithm.
	
	These prefactors were sequentially used to construct the single-mode-squeezing-like operator in Eq.~\eqref{Eq:App:Squeez:Single:Mode} by combining built-in functions in \texttt{Qutip} for defining creation and annihilation operators, with linear algebra functions of \texttt{Scipy} that allow the exponentiation of matrices. Given that \texttt{Qutip} expresses quantum optical states and operators in the Fock basis, which is inherently infinite dimensional, it is important to select a sufficiently high cutoff dimension for the Fock basis to ensure accurate representation of quantum states, while not exceeding the memory capabilities of the employed hardware. For the parameter regime we worked with, we found that $n_{\text{cutoff}} = 50$ was sufficient to achieve this balance, although convergence tests were performed with $n_{\text{cutoff}} = 200$. These conditions were used for evaluating the optimization problem in Eq.~\eqref{Eq:App:optimization}, which was addressed through brute-force search. This method is particularly useful here, as we have a single parameter $\theta \in [0,\pi]$. Consequently, we generated an array composed of 100 elements linearly distributed within this interval and proceed to sequentially evaluate Eq.~\eqref{Eq:App:optimization} for each harmonic mode individually.
	
	\section{Details about the two-mode squeezing analysis}\label{App:TWS}
	In the analysis of Sec.~\ref{App:SMS}, we have purposely omitted the contribution of correlations between different field modes. In other words, we have neglected cross terms of the form $\hat{a}_q\hat{a}_{q'}^\dagger$ with $q\neq q'$. However, these correlations are present and must be characterized. The main challenge here is that the number of harmonic modes in this problem is large, making it numerically daunting to characterize all modes simultaneously.~Consequently, we restrict our analysis to two-mode squeezing contributions affecting only modes $q_1$ and $q_2$.~As observed in Fig.~\ref{Fig:Eig:Covar}, this approximation is particularly valid as, among all the modes involved, the correlations are specially significant between modes $q=1$ and $3$, while for the others they are almost negligible.
	
	Provided that the states we are dealing with are Gaussian states, their properties are fully determined by their covariance matrix $\sigma_{q_1,q_2}$, defined as in Eq.~\eqref{Eq:Cov:matrix}.~For simplicity, in the numerical calculations of this matrix we have omitted the contribution of the displacement operator appearing in Eq.~\eqref{Eq:two:mode:approx}.~This omission is justified because such displacement operators do not affect the correlation properties of the covariance matrix; their influence can be easily removed by means of local unitary transformations~\cite{duan_inseparability_2000,horodecki_quantum_2009}.
	
	One of the most interesting aspects about Gaussian states is that the PPT or Peres-Horodecki criterion is a necessary and sufficient condition for separability for all bipartite Gaussian states~\cite{simon_peres-horodecki_2000}. Thus, the logarithmic negativity is an ideal entanglement measure to characterize the entanglement in these states. This quantity can be computed as~\cite{horodecki_quantum_2009}
	\begin{equation}
		E_N(\hat{\rho})
		= \max\big\{0, -\log_2(2 \Tilde{\nu}_-)\big\},
	\end{equation}
	where $\Tilde{\nu}_-$ is the smallest of the two symplectic eigenvalues of the partial transpose of our covariance matrix. For the case of Gaussian states, the partial transpose can be computed as $\Tilde{\sigma} = \Gamma \sigma \Gamma$ with $\Gamma = \text{diag}(1,1,1,-1)$~\cite{simon_peres-horodecki_2000}. The symplectic eigenvalues of $\Tilde{\sigma}$ are then computed as the eigenvalues of the matrix $|i\Omega \sigma|$ with $\Omega = \bigoplus_{i=1}^n \Lambda$ with $\Lambda = \mqty(0& 1 \\ -1 & 0)$~\cite{adesso_gaussian_2005}. However, when considering matrices of the form of Eq.~\eqref{Eq:Cov:matrix}, $\Tilde{\nu}_{-}$ can also be computed as~\cite{adesso_gaussian_2005}
	\begin{equation}
		\Tilde{\nu}_{-}
		= \sqrt{\dfrac{\Delta(\Tilde{\sigma}) - \sqrt{\Delta(\Tilde{\sigma})^2-4\det(\sigma)}}{2}},
	\end{equation}
	with $\Delta(\Tilde{\sigma}) = \det(A) + \det(B)-2\det(C)$, with $\det(\cdot)$ denoting the determinant operation.
	
	\begin{figure}[h!]
		\centering
		\includegraphics[width=0.7\textwidth]{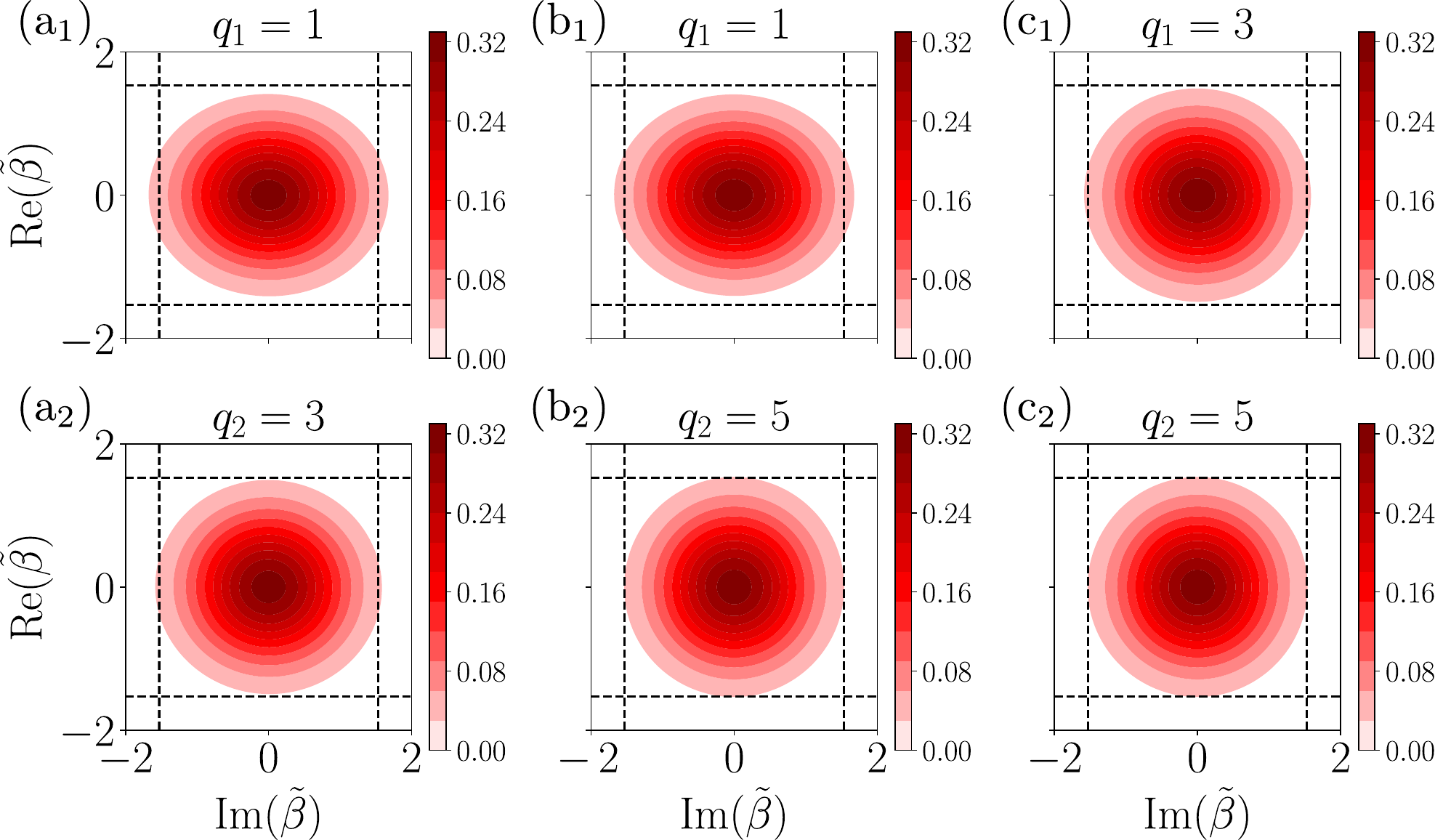}
		\caption{Wigner functions for the two-mode squeezing involving modes $q_1$ and $q_2$. The amount of squeezing found, in dB units, is (a$_1$) $6.71\times 10^{-2}$, (a$_2$) $8.46\times 10^{-3}$, (b$_1$) $6.71\times 10^{-2}$, (b$_2$) $4.63\times 10^{-14}$, (c$_1$) $8.46\times 10^{-3}$ and (c$_2$) $5.98\times 10^{-14}$.}
		\label{Fig:Wigners:prior}
	\end{figure}
	
	Similar to Sec.~\ref{App:SMS}, the prefactors of the operators $\{\hat{a}_{q_i}^2, \hat{a}_{q_i}^{\dagger 2} , \hat{a}_{q_i}\hat{a}_{q_i}^\dagger, \hat{a}_{q_i}^\dagger \hat{a}_{q_i}, \hat{a}_{q_i}\hat{a}_{q_j}^\dagger, \hat{a}_{q_i}^\dagger \hat{a}_{q_j}\}$ were numerically computed using the \texttt{quad} function of \texttt{Scipy}.~Then, the two-mode squeezing operator and the corresponding covariance matrix were computed by combining built-in \texttt{Qutip} functions with linear algebra functions of \texttt{Scipy}.~For both $q_1$ and $q_2$ modes, the upper bound used for their respective Hilbert space dimension was set to $n_{\text{cutoff}} = 50$, leading to a joint Hilbert space dimension of 2500.~Under these conditions, we obtained the Wigner functions in Fig.~\ref{Fig:Wigners:prior} for different combinations of $(q_1,q_2)$.
	
	The Wigner functions in Fig.~\ref{Fig:Wigners:prior} correspond to those of Fig.~\ref{Fig:Wigners:after} prior to the performance of the heralding operation. That such heralding operations can lead to other kinds of non-classical states of light becomes evident when expanding the two-mode-squeezing-like operator in Eq.~\eqref{Eq:two:mode:approx} in polynomial series.~By denoting, for instance, the prefactor of the operator $\hat{a}_{q_i}^2$ as $g(\hat{a}_{q_i}^2)$, and by considering only first-order terms while omitting the effect of the displacement operator, we can express our state as (up to normalization factors)
	\begin{equation}
		\begin{aligned}
			\lvert \bar{\Phi}_{\text{e}}(t) \rangle
			&\approx \ket{\bar{0}} 
			+ \sum^2_{i,j = 1}
			\big[
			g(\hat{a}_{q_i}\hat{a}_{q_i})
			\hat{a}_{q_i} \hat{a}_{q_j}
			+g(\hat{a}^\dagger_{q_i}\hat{a}^\dagger_{q_i})
			\hat{a}^\dagger_{q_i} \hat{a}^\dagger_{q_j}
			+g(\hat{a}^\dagger_{q_i}\hat{a}_{q_i})
			\hat{a}^\dagger_{q_i} \hat{a}_{q_j}
			+g(\hat{a}_{q_i}\hat{a}^\dagger_{q_i})
			\hat{a}_{q_i} \hat{a}^\dagger_{q_j}
			\big]\ket{\bar{0}}
			\\&=
			\ket{\bar{0}}
			+ \sum^2_{i,j = 1}
			g(\hat{a}^\dagger_{q_i}\hat{a}^\dagger_{q_j})
			\hat{a}^\dagger_{q_i}\hat{a}^\dagger_{q_j}
			\ket{\bar{0}},
		\end{aligned}
	\end{equation}
	which after performing the heralding operation, for instance in the second mode, leads to
	\begin{equation}
		\hat{\rho}_{q_1}
		= \text{tr}_{q_2}(\hat{\Pi}_{q_2} 	\lvert\bar{\Phi}_{\text{e}}(t)\rangle\!\langle\bar{\Phi}_{\text{e}}(t)\rvert)
		= \abs{g(\hat{a}_{q_2}^{\dagger 2})}^2 
		\dyad{0}
		+\abs{g(\hat{a}_{q_1}^{\dagger}\hat{a}_{q_2}^{\dagger})}^2
		\dyad{1},		 
	\end{equation}
	with the obtained non-classical features strongly dependent on the ratio between the single-mode squeezing found in mode $q_2$ ($\lvert g(\hat{a}_{q_2}^{\dagger 2})\rvert^2$), and the strength of the correlations between $q_1$ and $q_2$ ($\lvert g(\hat{a}_{q_1}^{\dagger}\hat{a}_{q_2}^{\dagger})\rvert^2$).~As observed in Fig.~\ref{Fig:Amount:Squeezing}, the single-mode squeezing properties of our state decrease for higher-harmonic orders, as well as the correlations between the modes, although with different scaling as observed in Fig.~\ref{Fig:Eig:Covar}.
	
	Thus, when performing the heralding on $q_2 > q_1$ (first row of Fig.~\ref{Fig:Wigners:after}), it leads to higher non-classical features because the single-mode squeezing properties in these higher-order modes are almost vanishing.~On the contrary, when performing the heralding in mode $q_1$ with $q_2 > q_1$ (second row of Fig.~\ref{Fig:Wigners:after}), the single-mode squeezing properties are dominant, and therefore the presence of non-classical features become more diluted (panel (a$_2$)), if existing at all (panels (b$_2$) and (c$_2$)).
	
\end{document}